\documentclass[twocolumn,nopacs,prb,amsfonts,amsmath,amssymb,floatfix]{revtex4} 
\usepackage{color}
\usepackage{hhline}	
\usepackage{mathrsfs}
\usepackage{graphicx}
\usepackage{dcolumn}
\usepackage{bm}
\usepackage{multirow}
\usepackage{booktabs}
\usepackage{afterpage}
\usepackage{amsmath}
\usepackage{braket}
\usepackage{here}

\begin{document}

\title{Thermoelectric performance of materials with Cu$Ch_4$ ($Ch=$ S, Se) tetrahedra: Similarities and differences among their low-dimensional electronic structure from first principles}

\author{Masayuki Ochi}
\author{Hitoshi Mori}
\author{Daichi Kato}
\author{Hidetomo Usui}
\author{Kazuhiko Kuroki}
\affiliation{Department of Physics, Osaka University, Machikaneyama-cho, Toyonaka, Osaka 560-0043, Japan}

\date{\today}
\begin{abstract}
In this study, we perform a comparative theoretical study on the thermoelectric performance of materials with Cu$Ch_4$ ($Ch=$ S, Se) tetrahedra, including famous thermoelectric materials BiCuSeO and tetrahedrite Cu$_{12}$Sb$_4$S$_{13}$, by means of first-principles calculations.
By comparing these electronic band structures, we find that many of these materials possess a Cu-$t_{2g}$ band structure consisting of quasi-one-dimensional band dispersions and the isotropic (two-dimensional for layered compounds) band dispersion near the valence-band edge.
Therefore, the key factors for the thermoelectric performance are the anisotropy of the former band dispersion and the degeneracy of these two kinds of band dispersions.
We also find that a large extension of the chalcogen orbitals often improves their thermoelectric performance by improving these two factors or by going beyond such a basic band structure through a large alternation of its shape. 
Such a large extension of the chalcogen orbitals might partially originate from the anisotropic Cu-$Ch$ bond geometry of a tetrahedron.
Our study reveals interesting similarities and differences of materials with Cu$Ch_4$, which provides important knowledge for a future search of high-performance thermoelectric materials.
\end{abstract}

\maketitle

\section{Introduction}

Exploring high-performance thermoelectric materials is one of the most important tasks for solving the energy problem. The efficiency of the thermoelectric conversion is governed by the dimensionless figure of merit $ZT=\sigma S^2 T \kappa^{-1}$, where $\sigma$, $S$, $T$, and $\kappa$ are the electrical conductivity, the Seebeck coefficient, the temperature, and the thermal conductivity, respectively. 
One promising and popular way to increase $ZT$ is by reducing the lattice thermal conductivity, e.g., by nanostructuring~\cite{nano1,nano2,nano3}.
Another way is to increase the power factor $\mathrm{PF}= \sigma S^2$, several kinds of desirable band structures for which have been proposed, such as a sharp peak of the density of states (DOS) near the Fermi level based on the Mott formula~\cite{Mott}, a pudding-mold-shaped band~\cite{pudding}, 
high degeneracy (band convergence) of the band edges (e.g.~Ref.~[\onlinecite{band_conv_PbCh}]), and low dimensionality~\cite{Hicksone,Hicksone2,Dress,usuione,Fukuyama}.
For example, a high-throughput search of thermoelectric materials was proposed using the Fermi surface complexity factor~\cite{FSCfactor}, which becomes larger when the valley degeneracy becomes higher and the anisotropy of the effective mass becomes stronger.
The entropy originating from the degeneracy of $3d$ electrons together with their strong correlation effects is also an intriguing source of a high thermopower~\cite{entropy}.

However, an ideal electronic structure is difficult to realize because of the high sensitivity of the power factor to the electronic band structure near its edge within a few $k_B T \sim$ an order of 10 or 100 meV.
Therefore, a theoretical investigation to control the band structure has usually been performed individually for a specific kind of materials, and so a general observation for a wider group of materials has still been missing, while it is rather important when searching for a new material from the knowledge of existing compounds.

\begin{figure}
\begin{center}
\includegraphics[width=8.4 cm]{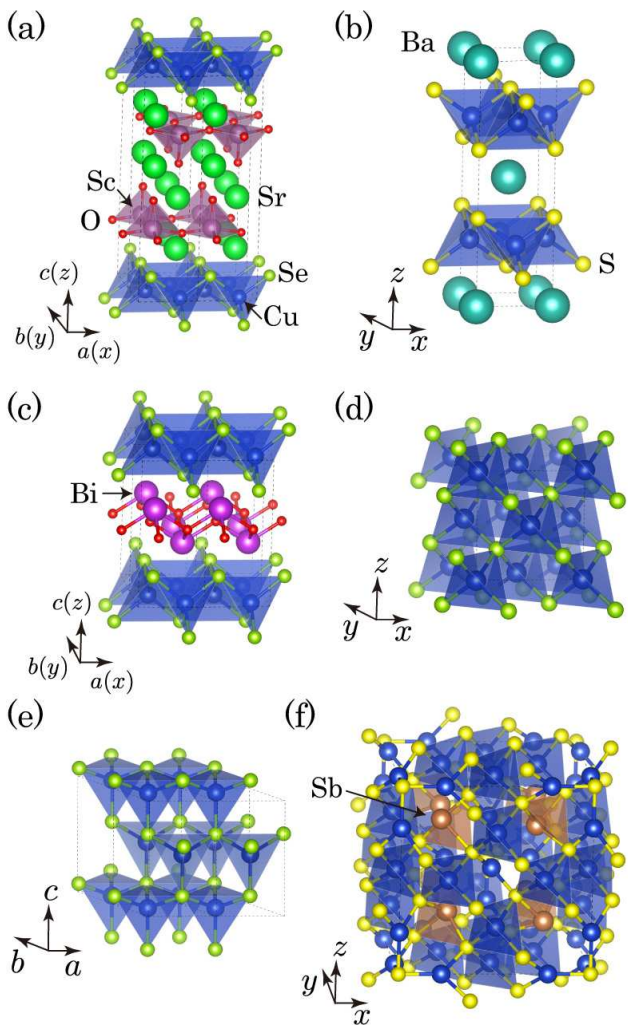}
\caption{Crystal structures of target materials in this study: (a) Sr$_2$ScCuSeO$_3$, (b) $\beta$-BaCu$_2$S$_2$, (c) BiCuSeO, (d) zincblende CuSe (hypothetical), (e) wurtzite CuSe (hypothetical), and (f) Cu$_{12}$Sb$_4$S$_{13}$. Cu$Ch_4$ tetrahedra are colored with blue.
These figures were depicted using the VESTA software~\cite{VESTA}.}
\label{fig:crystal}
\end{center}
\end{figure}

In this paper, we perform a comparative theoretical study on the thermoelectric performance of materials with Cu$Ch_4$ ($Ch=$ S, Se) tetrahedra, which can be found in famous thermoelectric materials such as BiCuSeO~\cite{BiCuSeO_org,BiCuSeO_highZT1,BiCuSeO_highZT2,BiCuSeO_review} and tetrahedrite Cu$_{12}$Sb$_4$S$_{13}$~\cite{tetrahed_1,tetrahed_2,tetrahed_3,tetrahed_4,tetrahed_5,tetrahed_6,tetrahed_7,tetrahed_review} (see Fig.~\ref{fig:crystal}), by means of first-principles calculations.
Through a careful investigation into their electronic structure, we reveal several similarities and differences among them.
Many of these materials have a valence-band structure consisting of quasi-one-dimensional band dispersions with (nearly) two-fold degeneracy at their band edges, together with the rather isotropic (or two-dimensional for layered structures) band dispersion.
Therefore, such a band structure can be basically characterized by the degeneracy of these three band edges~\cite{122_degen} and the anisotropy (quasi-one-dimensionality) of the band dispersion.
An intriguing characteristic of the electronic structure of the Cu$Ch_4$ tetrahedra is that chalcogen atomic orbitals spreading into the void space in the crystal often enhance the thermoelectric performance by improving the two key factors mentioned above or by going beyond such a basic band structure through a large alternation of its shape.
It is also characteristic that the band splitting induced by the spin-orbit coupling (SOC) in the non-centrosymmetric environment of the tetrahedron degrades the thermoelectric performance at low temperatures.
Our observation will be important to the search for new high-performance thermoelectric materials including tetrahedra.

This paper is organized as follows. In Sec.~\ref{sec:method}, we show the detail of the calculation methods we used in this study. 
Sections~\ref{sec:gen}, \ref{sec:122}, \ref{sec:BiCuSeO}, \ref{sec:ZBWZ}, and \ref{sec:tetra} present calculation results for Sr$_2$ScCu$Ch$O$_3$~\cite{Sr2ScCuSO3,Sr2ScCuSeO3}, $\beta$-BaCu$_2$S$_2$~\cite{BaCu2S2_1,BaCu2S2_2,BaCu2S2_3,BaCu2S2_4,BaCu2S2_5,BaCu2S2_6,BaCu2S2_7}, BiCuSeO, Cu$Ch$ with hypothetical zincblende and wurtzite structures, and Cu$_{12}$Sb$_4$S$_{13}$, respectively.
In Sec.~\ref{sec:gen}, we also present some general perspectives on materials with Cu$Ch_4$ tetrahedra.
Some materials with Cu$Ch_4$ tetrahedra that are not investigated in our study are briefly discussed in Sec.~\ref{sec:others}.
Section~\ref{sec:sum} is devoted to a summary of this study.

\section{Methods of calculations\label{sec:method}}

First, we determined the crystal structures through structural optimization using the Perdew--Burke--Ernzerhof parametrization of the generalized gradient approximation (PBE-GGA)~\cite{PBE} and the projector augmented wave method~\cite{paw} as implemented in the \textsc{VASP} code~\cite{vasp1,vasp2,vasp3,vasp4} except BiCuSeO, for which we employed the experimental structure taken from Ref.~[\onlinecite{BiCuSeO_expt}]~\cite{note_BiCuSeO}.
Plane-wave cutoff energies for the structural optimization were 550 eV for Sr$_2$ScCu$Ch$O$_3$ and 400 eV for other materials.
The $\bm{k}$-meshes used in the structural optimization were $10\times10\times 4$ for Sr$_2$ScCu$Ch$O$_3$, $8\times8\times8$ for $\beta$-BaCu$_2$S$_2$ and Cu$_{12}$Sb$_4$S$_{13}$, and $16\times16\times16$ for other materials.
For the structural optimization, we always included the SOC, while the band-structure calculations with and without SOC were both performed using those structures.
For the first-principles band-structure calculation, we used PBE-GGA and the full-potential (linearized) augmented plane-wave method, as implemented in the \textsc{wien2k} code~\cite{wien2k}. The $RK_{\rm max}$ parameter was set to 8.0.

After the first-principles band-structure calculation, we extracted the Wannier functions from the calculated band structures using the \textsc{wien2wannier} and \textsc{wannier90} codes~\cite{Wannier1,Wannier2,Wannier90,Wien2Wannier}.
In this study, we took the following orbitals as the Wannier basis set: Cu-$d$, $Ch$-$p$, and O-$p$ orbitals for Sr$_2$ScCu$Ch$O$_3$, 
Cu-$d$ and $Ch$-$p$ orbitals for $\beta$-BaCu$_2$S$_2$ and Cu$Ch$ with hypothetical zincblende and wurtzite structures, 
Cu-$d$ orbitals for Cu$_{12}$Sb$_4$S$_{13}$, 
and Cu-$s,d$, Se-$p$, and Bi-$p$ orbitals for BiCuSeO.
For BiCuSeO, the outer and inner windows for Wannierization were set to [$-8$:$8$] and [$-8$:$3$] eV, respectively, where the valence-band top was set to zero. For other materials, band structures with this Wannier basis set are isolated (i.e. not entangled), so that there are no degrees of freedom to adjust the windows.
We did not perform the maximal localization procedure for constructing the Wannier functions to prevent orbital mixing among the different spin components and to allow for a more intuitive understanding of the hopping parameters. While this issue is only related to calculations with SOC, we did not perform the maximal localization also for calculations without SOC in order to describe the electronic structure on an equal footing for both cases.
We constructed the tight-binding model with the obtained hopping parameters among the Wannier functions, and we analyzed the transport properties using this model.
For this purpose, we employed Boltzmann transport theory, where the transport coefficients ${\bf K}_{\nu}$ are represented as follows:
\begin{align}
{\bf K}_{\nu}= \tau\int \mathrm{d}\bm{k} \sum_{n} \bm{v}_{n,\bm{k}}\otimes\bm{v}_{n,\bm{k}}\left[-\frac{\partial f_0}{\partial \epsilon_{n,\bm{k}}}\right](\epsilon_{n,\bm{k}}-\mu(T))^\nu ,\label{eq:transp}
\end{align}
by using the Fermi--Dirac distribution function $f_0$, the chemical potential $\mu(T)$, the energy $\epsilon_{n,\bm{k}}$ and the group velocity $\bm{v}_{n,\bm{k}}$ of the one-electron orbital on the $n$-th band at some $\bm{k}$-point, and the relaxation time $\tau$, which was assumed to be constant ($\tau=10^{-15}$ second) in this study.
In reality, the relaxation time depends on many parameters such as temperature, $\bm{k}$-point, band index, energy, and direction. A different band structure generally yields a different relaxation time, and several well-known characteristics such as the band degeneracy and low dimensionality, which are considered to be favorable for high power factor, can shorten the relaxation time by the increased scattering rate (such as that by phonons) and somewhat weaken their superiority. Nevertheless, our investigations into the band structures still offer important knowledge for understanding the thermoelectric performance of our target materials. In other words, our main objective in this study is to compare the shape of the band structure and discuss how to realize a favorable one in materials with Cu$Ch_4$ tetrahedra.
Although the material dependence of $\tau$ is an important issue, it is beyond the scope of this study.
Here, $\mu(T)$ was determined to provide a given carrier density against the temperature change for calculations with a fixed carrier density.
The electrical conductivity ${\boldsymbol \sigma}$ and the Seebeck coefficient ${\bf S}$ are expressed as follows:
\begin{align}
{\boldsymbol \sigma}=e^2{\bf K}_0,\ \ \  {\bf S}=-\frac{1}{eT}{\bf K}_0^{-1}{\bf K}_1,
\end{align}
where $e$ ($>0$) is the elementary charge.
The power factor PF $=\sigma S^2$ was calculated using the diagonal components of these tensors.
For layered materials (Sr$_2$ScCu$Ch$O$_3$, $\beta$-BaCu$_2$S$_2$, and BiCuSeO), we only calculated the transport quantities along the $x$ direction, because the $y$ direction is equivalent to the $x$ direction and the conductivity along the $z$ direction is much smaller.
Also for Cu$Ch$ with the hypothetical zincblende structure and Cu$_{12}$Sb$_4$S$_{13}$, we only showed the transport quantities along the $x$ direction because of the isotropy, i.e., the $y$ and $z$ directions are equivalent to the $x$ direction.
For Cu$Ch$ with the hypothetical wurtzite structure, we showed the transport quantities along the $x$ and $z$ directions.
To simulate the carrier doping, we adopted the rigid band approximation.
In this study, we investigated the transport properties at $T=300$ K unless noted otherwise.
For transport calculations using the Boltzmann transport theory described above, we ignored the contribution from conduction bands to concentrate on the transport properties of the valence-band structure.
For evaluating the effective mass, we fitted a band dispersion with a parabola function within a normalized displacement in the Brillouin zone $\Delta k_i = \pi/20$ ($i=x,y,z$) from the valence-band top.

\section{Results and Discussion\label{sec:res}}

\subsection{Sr$_2$ScCu$Ch$O$_3$ and general perspectives on materials with Cu$Ch_4$ tetrahedra\label{sec:gen}}

We first investigated the electronic structure of Sr$_2$ScCu$Ch$O$_3$ ($Ch=$ S, Se) where Cu$Ch_4$ tetrahedra constitute a layered structure (Cu$Ch$ layers) separated by insulating perovskite layers consisting of the Sr, Sc, and O atoms.
These compounds are kinds of mixed-anion compounds~\cite{mixed_review}.
Because some of their analogous compounds are known to be p-type in experiments owing to Cu vacancies~\cite{pero_hole1,pero_hole2}, we concentrated on the hole carrier doping (i.e. the valence-band structure).
Because the valence-band structure near its edge is governed by the nearly isolated Cu$Ch$ layers with high symmetry, Sr$_2$ScCu$Ch$O$_3$ can be regarded as a prototypical system to investigate basic properties existing in materials with Cu$Ch_4$ tetrahedra.
Therefore, the main purpose of this subsection is to extract general perspectives on materials with Cu$Ch_4$ tetrahedra from the analysis on Sr$_2$ScCu$Ch$O$_3$.

\subsubsection{Band structure and its characteristics}

\begin{figure}
\begin{center}
\includegraphics[width=8 cm]{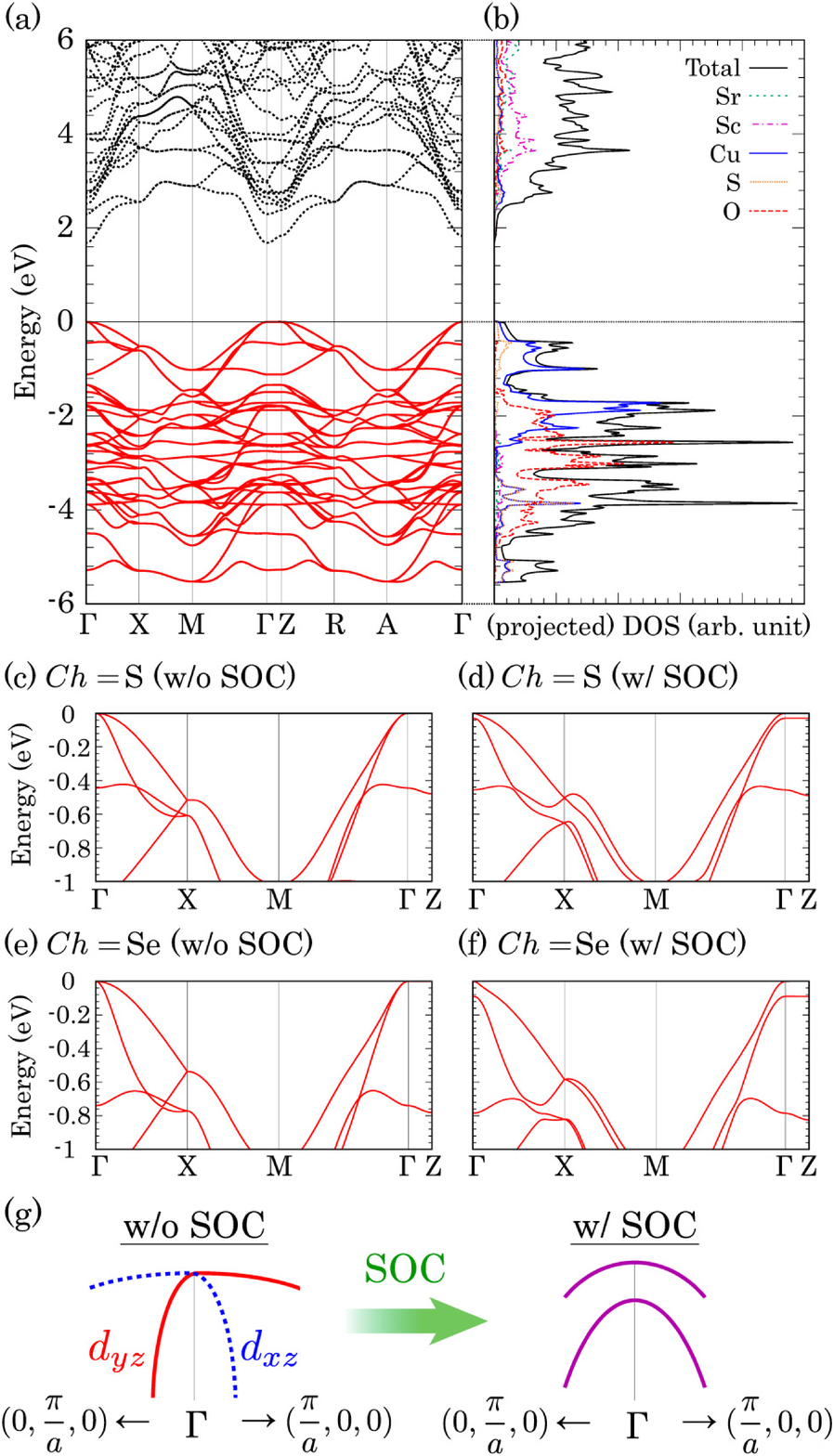}
\caption{(a) Band structure and (b) (projected) DOS of Sr$_2$ScCuSO$_3$ calculated without SOC. Black broken and red solid lines in (a) represent the band structures obtained with the first-principles calculation and the tight-binding model of the Wannier functions, respectively. (c)--(f) Blow-up views of the first-principles band structures near the valence-band top for Sr$_2$ScCu$Ch$O$_3$ ($Ch=$ S, Se) with and without SOC. (g) Schematic picture for the impact of SOC on the $d_{xz/yz}$ band dispersions near the valence-band top at the $\Gamma$ point.}
\label{fig:band_pero}
\end{center}
\end{figure}

Figure~\ref{fig:band_pero}(a)(b) shows the band structure and (projected) DOS of Sr$_2$ScCuSO$_3$ calculated without SOC.
Red solid lines in Fig.~\ref{fig:band_pero}(a) show the band structure obtained with the tight-binding model of the Wannier functions constructed from the first-principles band structure shown with the black broken lines. We can see that the valence-band top mainly consists of the Cu states strongly hybridized with the S states. 
While we denote the valence-band structures, as the `$d_{xz}$ band' in this paper, we note that the chalcogen $p$-orbitals are always strongly hybridized with them, and so this notation is used just for simplicity.
Because of the existence of the insulating layers, the conduction along the $z$ direction is almost prohibited, at least near the valence-band top as inferred from the flat band dispersion along the $\Gamma$-Z line shown in Fig.~\ref{fig:band_pero}(a).

Blow-up views of the first-principles band structures near the valence-band top are shown in Fig.~\ref{fig:band_pero}(c)--(f) for $Ch=$ S, Se with and without SOC. For all these figures, the Cu-$d_{xz/yz}$ bands exist near the valence-band top of the $\Gamma$ point.
A difference in the group velocity (or the effective mass) for these two band dispersions along the $\Gamma$-X line corresponds to the anisotropy of the Cu-$d_{xz/yz}$ bands. In other words, along the $x$ direction, the Cu-$d_{xz}$ band should be more dispersive than the Cu-$d_{yz}$ band by the orbital anisotropy, and vice versa for the $y$ direction. This situation is shown schematically in the left half of  Fig.~\ref{fig:band_pero}(g).
Therefore, the valence-band structure near its edge can be regarded as quasi-one-dimensional.
Such a low dimensionality is a key aspect for obtaining a good thermoelectric performance because it increases the DOS near the band edge and therefore the PF~\cite{Hicksone,Hicksone2,Dress,usuione}.
Note that, when one varies the effective mass of the isotropic band dispersion, an increase of the DOS lowers the group velocity and vice versa; such a trade-off relation 
makes it difficult to maximize the thermoelectric performance through band engineering. On the other hand, introducing the low dimensionality increases the DOS without degrading the group velocity with respect to a specific direction, which is why the low dimensionality is regarded as one of the ideal band structures for the thermoelectric performance.
A Cu-$d_{x^2-y^2}$ band lies around 0.4 eV below the valence-band top for $Ch=$ S as shown in Fig.~\ref{fig:band_pero}(c)--(d), while this energy distance is almost doubled for $Ch=$ Se as shown in Fig.~\ref{fig:band_pero}(e)--(f).
Note that the $t_{2g}$ orbitals correspond to the $d_{xz/yz/x^2-y^2}$ orbitals in the present coordinate.
Here, a triple degeneracy of the $t_{2g}$ orbitals is lifted in the crystal field of this material, and so these orbitals should not be called the $t_{2g}$ orbitals in the strict sense of the term. However, for simplicity and convenience for comparing the band structures consisting of these orbitals in other materials with different crystal symmetries as we shall investigate, we simply call them the `$t_{2g}$' orbitals in this paper.

Another important feature of the band structure is a non-negligible band splitting just near the valence-band top in Fig.~\ref{fig:band_pero}(f), which is induced by SOC, while
the Cu-$d_{xz/yz}$ dispersions are degenerate at the valence-band top if SOC is switched off as is expected from the crystal symmetry (i.e. the equivalence between the $x$ and $y$ directions).
The effect of SOC is schematically shown in Fig.~\ref{fig:band_pero}(g).
Quasi-one-dimensional Cu-$d_{xz/yz}$ bands are hybridized by SOC, and then they become two isotropic two-dimensional bands with a small gap.
The size of this gap is around 30 and 90 meV for $Ch=$ S and Se, respectively.
This band hybridization is pronounced only in the region near the $\Gamma$ point where the $d_{xz/yz}$ bands become close, which we can verify by comparing Fig.~\ref{fig:band_pero}(e) and (f). Nevertheless, because the valence-band-top structure is altered by SOC, the power factor is affected as we shall see later in this paper.

\begin{table}
\begin{center}
\begin{tabular}{c c c c}
\hline \hline
 $Ch$& & w/o SOC & w/ SOC \\
\hline
S & Highest valence band & 2.94 & 0.97 \\  
   & Next highest valence band & 0.47 & 0.71 \\
   & Mass anisotropy $\gamma$ & 6.2 & -\\
\hline
Se & Highest valence band & 2.91 & 0.57 \\  
     &  Next highest valence band & 0.32 & 0.59 \\
        & Mass anisotropy $\gamma$ & 9.0 & -\\
\hline \hline
\end{tabular}
\caption{The effective masses along the $x$ direction, $m^*_{xx}/m_e$, of the first and second highest valence bands for Sr$_2$ScCu$Ch$O$_3$ ($Ch=$ S, Se). The anisotropy of the effective mass $\gamma$, which corresponds to the ratio of $m^*_{xx}$ for these two bands, is also shown for the band structure without SOC.\label{tab:pero}}
\end{center}
\end{table}

Table~\ref{tab:pero} presents the effective masses along the $x$ direction, $m^*_{xx}/m_e$, of the first and second highest valence bands, where $m_e$ is the free-electron mass.
In the case without SOC, a ratio of the effective masses $m^*_{xx}$ between the highest and next highest valence bands corresponds to the anisotropy of the effective mass for each $d_{xz/yz}$ band with respect to the $x$ and $y$ directions, as is explained above.
We call this ratio the mass anisotropy $\gamma$ hereafter, which is also shown in Table~\ref{tab:pero}.
In the case with SOC, we do not show the mass anisotropy because the valence bands near the $\Gamma$ point become isotropic in the $xy$ plane, as described above.
The most important thing shown in this table is the difference of the anisotropy between $Ch=$ S and Se, which we shall investigate in more detail in Sec.~\ref{sec:atom}.

\subsubsection{Transfer integrals and their roles\label{sec:hop}}

\begin{figure}
\begin{center}
\includegraphics[width=7.5 cm]{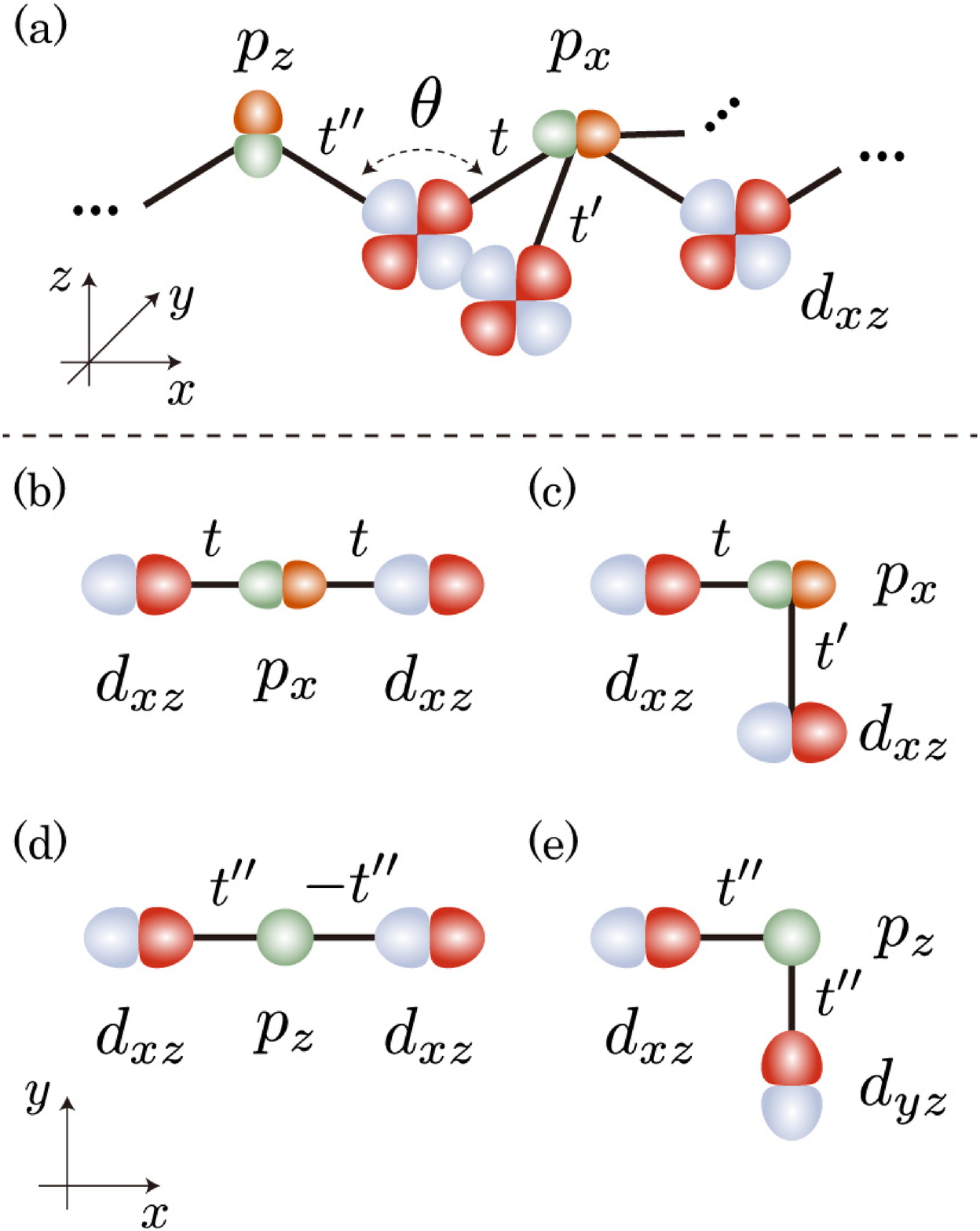}
\caption{(a) Definition of the nearest-neighbor transfer integrals between the Cu-$d_{xz/yz}$ and the $Ch$-$p$ orbitals: $t$, $t'$, and $t''$. The definition of the $Ch$-Cu-$Ch$ angle $\theta$ is also shown here. (b)--(e) Several hopping processes that can enhance or degrade the one-dimensionality of the $d_{xz/yz}$ bands.}
\label{fig:hop}
\end{center}
\end{figure}

\begin{figure}
\begin{center}
\includegraphics[width=8.5 cm]{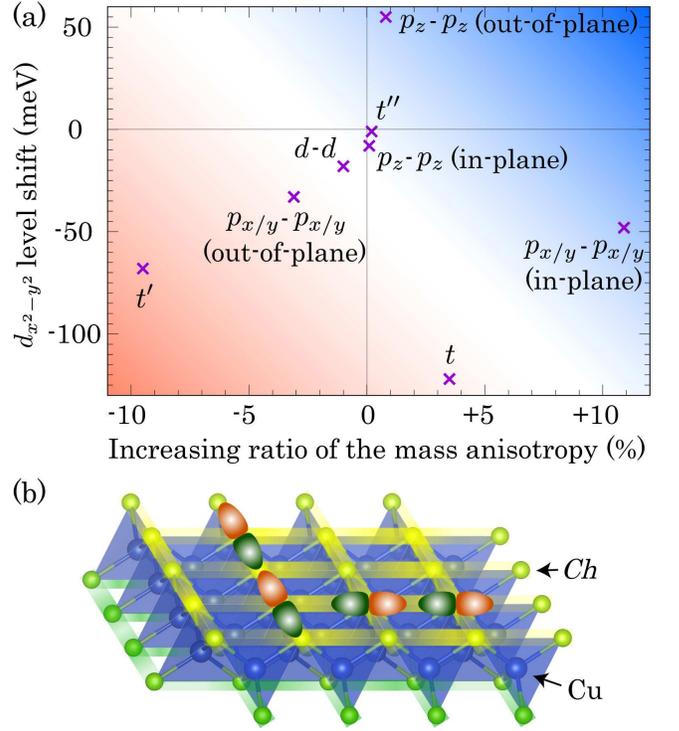}
\caption{(a) Plots for the shift of the $d_{x^2-y^2}$ level relative to the $d_{xz/yz}$ level at the $\Gamma$ point and the increasing ratio of the mass anisotropy when each (group of the) transfer integral(s) is increased by 10\% for Sr$_2$ScCuSO$_3$. SOC is switched off here in order to evaluate the mass anisotropy. (b) Schematic picture of the hidden chains on the $Ch$ square lattices in the Cu$Ch$ layer.}
\label{fig:hop_change}
\end{center}
\end{figure}

First, we analyzed microscopic origins that determine the valence-band structure near the band edge of Sr$_2$ScCu$Ch$O$_3$.
For this purpose, we investigated which transfer integrals in the tight-binding model for the Wannier functions contribute to the valence $t_{2g}$ band structure.

We begin with the definition of all the inequivalent nearest-neighbor transfer integrals between the Cu-$d_{xz/yz}$ and the $Ch$-$p$ orbitals: $t$, $t'$, and $t''$ as shown in Fig.~\ref{fig:hop}(a).
Figures~\ref{fig:hop}(b)(d) and (c)(e) show possible hopping processes that might enhance or degrade the one-dimensionality of the $d_{xz/yz}$ bands, respectively, from a top view. Because the other hopping processes are also possible by considering many inequivalent $d$-$d$ and $p$-$p$ transfer integrals, the situation is not so simple. To obtain a clear insight, we hypothetically increased a specific portion of the transfer integrals by 10\% in the tight-binding model of Sr$_2$ScCuSO$_3$, and we evaluated how much the mass anisotropy $\gamma$ and the $d_{x^2-y^2}$ level relative to the $d_{xz/yz}$ level at the $\Gamma$ point changes, which is shown in Fig.~\ref{fig:hop_change}(a). SOC was switched off here in order to calculate the mass anisotropy. The notation `in-plane' and `out-of-plane' for the $p$-$p$ transfer integrals means whether the $c$ ($z$) coordinates of the two S atoms between which the transfer integral is considered are the same or not~\cite{footnote_inplane}.

Looking at Fig.~\ref{fig:hop_change}(a), we first notice that $t''$ has less of an effect on both the anisotropy of the $d_{xz/yz}$ bands and the $d_{x^2-y^2}$ level shift. This is because the nearest-neighbor $d_{xz/yz}$-$p_z$ hoppings become inactive at the $\Gamma$ point due to the Bloch phase. We can see this situation in Fig.~\ref{fig:hop}(d): when $k_x=0$, the same Bloch phase of the $d_{xz}$ orbitals on two sites makes two $d_{xz}$-$p_z$ transfers cancel with each other. Therefore, on the $k_x=0$ ($k_y=0$) line, the $p_z$ orbitals are decoupled with the $d_{xz}$ ($d_{yz}$) orbitals~\cite{footnote_sym}. In particular, the hopping process shown in Fig.~\ref{fig:hop}(e), which requires the coupling of the $p_z$ orbital with both the $d_{xz/yz}$ orbitals, is thus completely inactivated on the $\Gamma$-X line where the mass anisotropy is evaluated.
Whereas the hopping process shown in Fig.~\ref{fig:hop}(d) begins to be activated when one moves from the $\Gamma$ point by increasing $k_x$, its effect seems very small in Fig.~\ref{fig:hop_change}(a).

Comparing $t$ and $t'$ in Fig.~\ref{fig:hop_change}(a), we can see that the anisotropy is affected much more by $t'$. A possible explanation for this might be that $t$ contributes to two processes [shown in Fig.~\ref{fig:hop}(b)(c)] that enhance and degrade the anisotropy, while $t'$ only contributes to the process shown in Fig.~\ref{fig:hop}(c), which degrades the anisotropy. Therefore, it is desirable to reduce $t'$ to improve the one-dimensionality.

In Fig.~\ref{fig:hop_change}(a), we can also see that, even if one increases all the $d$-$d$ direct hopping paths, its effect is small on the valence-band structure near the $\Gamma$ point. On the other hand, $p$-$p$ hopping amplitudes have a large impact on the valence-band structure.
In particular, the in-plane $p_{x/y}$-$p_{x/y}$ hopping amplitudes play a very important role in enhancing the one-dimensionality of the $d_{xz/yz}$ bands.
Because there are {\it hidden chains} on the $Ch$ square lattices in the Cu$Ch$ layer~\cite{note_ladder}, as shown in Fig.~\ref{fig:hop_change}(b), it is natural to consider that the one-dimensional character of the $d_{xz/yz}$ bands is strongly assisted by the anisotropy of the $Ch$-$p_{x/y}$ orbitals, which have strong $\sigma$-bonds with respect to the $x/y$ directions, respectively.
This mechanism bears a remarkable resemblance to the mechanism in BiS$_2$ superconductors, where some of the authors recently predicted that the PF can be drastically increased when the one-dimensionality of the $p$-states on the square lattice is further enhanced by atomic substitution~\cite{BiS2_Usui, BiS2_thermo_ochi}.
The situation in which the inter-chalcogen bonds play an important role for materials properties also reminds us of the intriguing roles of the chemical bonds between arsenic in iron-based superconductors~\cite{Nohara_As}.
It was pointed out theoretically in SnSe that Sn-Sn hopping strongly enhances the one-dimensionality of the electronic structure and thus can further increase its power factor~\cite{Mori_SnSe}.

The above discussion teaches us which hopping amplitudes we should pay attention to when one tries to control the band structure of this material.
We shall consider how the valence-band structure changes by the substitution of the chalcogen atoms and the hypothetical variation of the $Ch$-Cu-$Ch$ angle in Sections~\ref{sec:atom} and \ref{sec:angle}, respectively.

\subsubsection{$Ch=$ S vs. Se\label{sec:atom}}

In Table~\ref{tab:pero}, we can see that the mass anisotropy $\gamma$ for $Ch=$ Se, 9.0, is much larger than that for $Ch=$ S, 6.2. 
As a matter of fact, the calculated peak value of the power factor, which we will call PF$_{\mathrm{max}}$ hereafter, is larger for $Ch=$ Se than that for $Ch=$ S; PF$_{\mathrm{max}}$ are 1.40 and 1.55 $\mu$Wcm$^{-1}$K$^{-2}$ for $Ch=$ S and Se, respectively, when SOC is switched off.

\begin{table}
\begin{center}
\begin{tabular}{c c c c c}
\hline \hline
 $Ch$& $|t|$ & $|t'|$ & $|t''|$ & $|t_{p,p;\sigma}|$ \\
\hline
S & 0.62 & 0.34 & 0.11 & 0.44 \\  
Se & 0.57 & 0.33 & 0.16 &  0.51 \\  
\hline \hline
\end{tabular}
\caption{Some model parameters extracted from the first-principles band structure calculation for Sr$_2$ScCu$Ch$O$_3$ ($Ch=$ S, Se) without SOC. Definition of $t$, $t'$, and $t''$ are shown in Fig.~\ref{fig:hop}(a).
$t_{p,p;\sigma}$
is the transfer integrals between the $Ch$-$p_x$ orbitals with the same $z$ coordinates and the neighboring $x$ coordinates. All in eV.\label{tab:hop_S_Se}}
\end{center}
\end{table}

To understand this difference, we compared some transfer integrals as listed in Table~\ref{tab:hop_S_Se}.
When looking into the $d$-$p$ hopping amplitudes $|t|$, $|t'|$, and $|t''|$, it is hard to understand that the PF value is enhanced for $Ch=$ Se.
This is because a smaller $|t|$ in $Ch=$ Se will worsen the one-dimensionality of the $d_{xz/yz}$ band dispersions as we have seen in Sec.~\ref{sec:hop}. 
In addition, $|t'|$ is almost the same between $Ch=$ S and Se, and a difference in $|t''|$ will have a small effect on the anisotropy of the electronic structure.
On the other hand, it is rather plausible that the enhanced one-dimensionality in $Ch=$ Se is largely due to the enhanced $p$-$p$ hopping amplitude, especially the in-plane $p_{x/y}$-$p_{x/y}$ ones, which were shown in Sec.~\ref{sec:hop} to be effective in increasing the anisotropy of the electronic structure.
Because the Se orbitals spread in space more widely than the S orbitals do, it is natural that the $p$-$p$ hopping amplitudes become larger in $Ch=$ Se than $Ch=$ S. A decrease of the $Ch$-Cu-$Ch$ angle $\theta$ from 113$^\circ$ to 108.2$^\circ$ might also play some role in increasing the in-plane $p_{x/y}$-$p_{x/y}$ hopping amplitudes, while it is a rather minor effect on the anisotropy as we shall see in the next section.

\begin{figure}
\begin{center}
\includegraphics[width=8.5 cm]{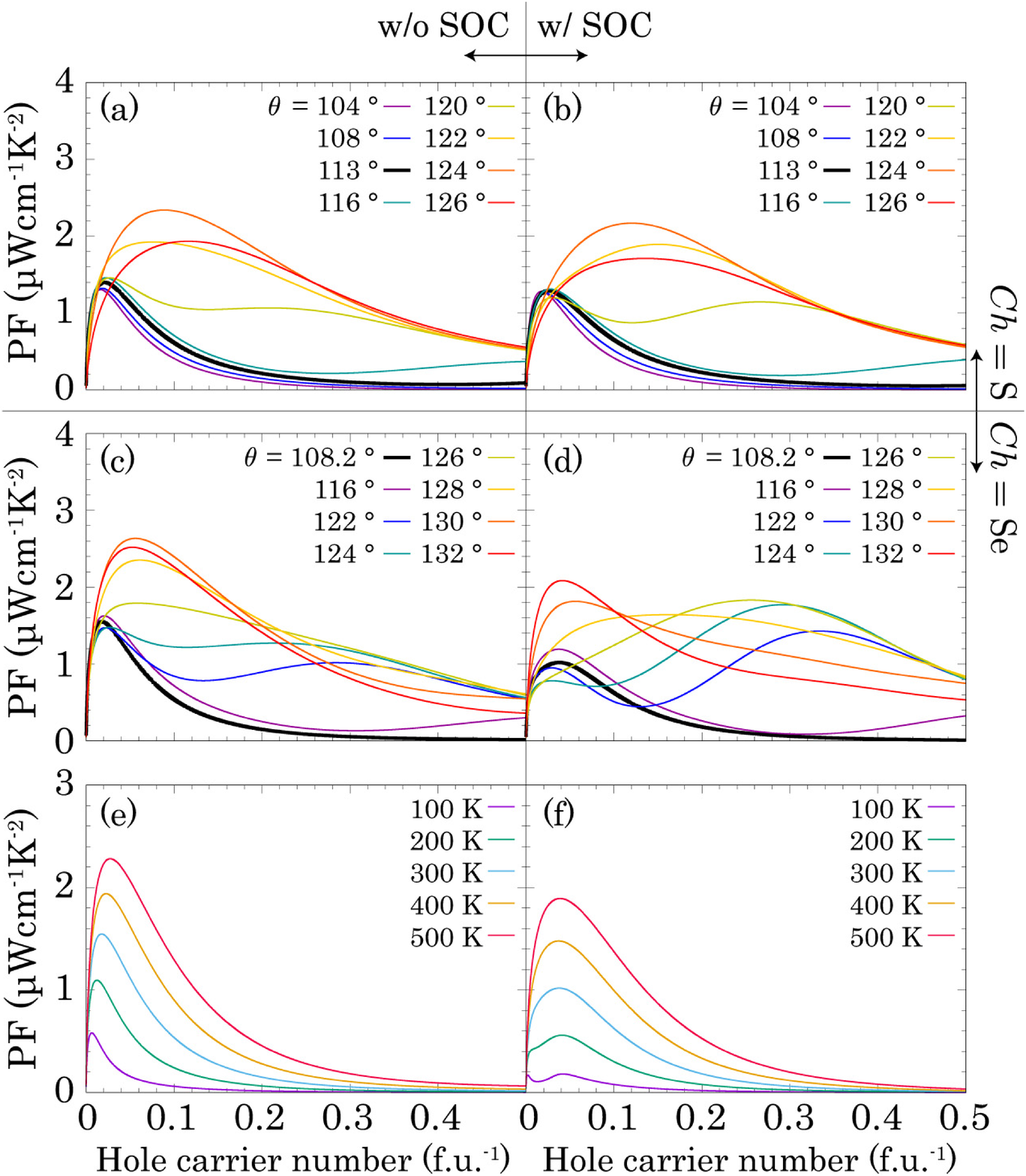}
\caption{Calculated PF values of Sr$_2$ScCu$Ch$O$_3$ with respect to the hole carrier number for (a) $Ch=$ S without SOC, (b) $Ch=$ S with SOC, (c) $Ch=$ Se without SOC, and (d) $Ch=$ Se with SOC, respectively. The $Ch$-Cu-$Ch$ angle $\theta$ as defined in Fig.~\ref{fig:hop}(a) is varied in the way described in Sec.~\ref{sec:angle}, where the original angles are 113$^\circ$ and 108.2$^\circ$ for $Ch=$ S and S, respectively. The PF curve for the original angle is shown with bold black lines in each panel. 
(e)(f) The same plots for $Ch=$ Se and the original bond angle $\theta=$ 108.2$^\circ$ using several temperatures obtained without and with SOC.
The relaxation time $\tau=10^{-15}$ second was used for all the panels.}
\label{fig:PF_angle}
\end{center}
\end{figure}

Whereas the anisotropy is enhanced for $Ch=$ Se, stronger SOC can degrade the thermoelectric performance.
As a matter of fact, PF$_{\mathrm{max}}$ exhibits a sizable reduction for $Ch=$ Se by introducing SOC while this effect is negligible for $Ch=$ S, as shown in Fig.~\ref{fig:PF_angle}(a)--(d), where black solid lines show PF curves for the original $Ch$-Cu-$Ch$ angle $\theta$.
The effect of the SOC hybridization is quite pronounced in low temperatures such as $T=$ 100 K as shown in Fig.~\ref{fig:PF_angle}(e)--(f), where the PF values calculated with and without SOC for $Ch=$ Se with several temperatures are shown.
For example, the PF peak shown in Fig.~\ref{fig:PF_angle}(f) splits into two for $T=$ 100 K.
The resulting PF$_{\mathrm{max}}$ calculated with SOC is around three times as small as that calculated without SOC for $T=$ 100 K.
This is in contrast to a relatively small impact on PF$_{\mathrm{max}}$ at higher temperatures such as 500 K, where the SOC gap of approximately 90 meV becomes less effective on transport properties by temperature effects.

Observation made in this section is summarized as follows. The enhanced one-dimensionality of the $d_{xz/yz}$ bands and the stronger SOC exist simultaneously for $Ch=$ Se. The stronger one-dimensionality is due to the larger hopping amplitudes between the in-plane $Ch$-$p_{x/y}$ orbitals on the hidden chains. However, PF$_{\mathrm{max}}$ calculated with SOC are 1.29 and 1.02 $\mu$Wcm$^{-1}$K$^{-2}$ for $Ch=$ S and Se, respectively, which means that the increased SOC hybridization has a larger impact on the power factor at $T=300$ K than the enhanced one-dimensionality. 

\subsubsection{Hypothetical variation of the $Ch$-Cu-$Ch$ angle\label{sec:angle}}

A Cu$Ch$ monolayer has only two degrees of freedom in its structure: the Cu-$Ch$ length and the $Ch$-Cu-$Ch$ angle $\theta$ as defined in Fig.~\ref{fig:hop}(a).
Here, we hypothetically vary the $Ch$-Cu-$Ch$ angle $\theta$ while fixing the Cu-$Ch$ length for Sr$_2$ScCu$Ch$O$_3$.
Concretely, the $c$ coordinate of the $Ch$ atoms and the lattice constant with respect to the $a$ and $b$ axes are changed while keeping the Cu-$Ch$ length unchanged from the original structure, while other atomic coordinates and the lattice constant with respect to the $c$ axis are fixed.

 \begin{figure}
\begin{center}
\includegraphics[width=8.5 cm]{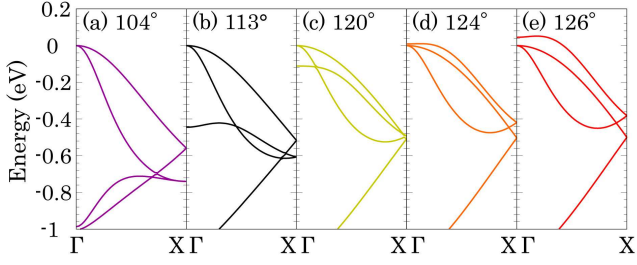}
\caption{The band structures of Sr$_2$ScCuSO$_3$ calculated without SOC are shown for several S-Cu-S angles $\theta$. Colors correspond to those shown in Fig.~\ref{fig:PF_angle}(a).}
\label{fig:band_angle}
\end{center}
\end{figure}

Figure~\ref{fig:PF_angle}(a)--(d) present the calculated PF with respect to the hole carrier number for $Ch=$ S without SOC, $Ch=$ S with SOC, $Ch=$ Se without SOC, and $Ch=$ Se with SOC, respectively. 
When one varies the $Ch$-Cu-$Ch$ angle by a few degrees from the original angles, we cannot see a large variation of the calculated PF.
However, PF peaks become twice as high as the original value for all the cases by a large increase of the angle.
This increase is induced by the degeneracy of the $d_{x^2-y^2}$ band edge with the $d_{xz/yz}$ band edge at the $\Gamma$ point.
In Fig.~\ref{fig:band_angle}(a)--(e), the valence-band structures are shown for several angles in the case of $Ch=$ S without SOC.
As is clearly seen, the $d_{x^2-y^2}$ level gets higher by increasing the $Ch$-Cu-$Ch$ angle, and when the PF peak is the largest value in Fig.~\ref{fig:PF_angle}(a), say, when the $Ch$-Cu-$Ch$ angle is 124$^\circ$, the three band edges are almost degenerate at the valence-band top as shown in Fig.~\ref{fig:band_angle}(d). A further increase degrades the PF peak value as shown in Fig.~\ref{fig:PF_angle}(a).
Corresponding to this observation, we can see that two PF peaks, e.g., $\theta=122^{\circ}$ in Fig.~\ref{fig:PF_angle}(c), get together by increasing the $Ch$-Cu-$Ch$ angle $\theta$.

\begin{figure}
\begin{center}
\includegraphics[width=6 cm]{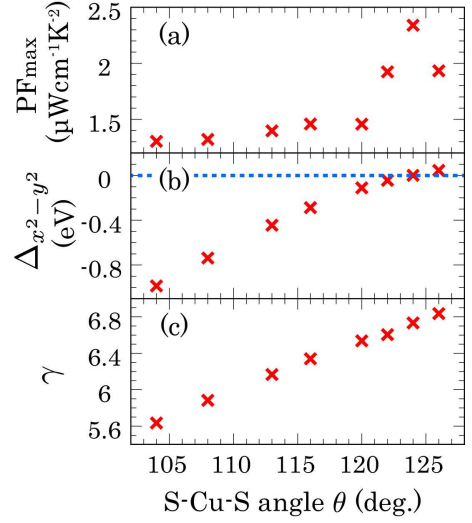}
\caption{(a) PF$_{\mathrm{max}}$, (b) the $d_{x^2-y^2}$ level relative to the $d_{xz/yz}$ level at the $\Gamma$ point, $\Delta_{x^2-y^2}$, and (c) the mass anisotropy $\gamma$ with respect to the S-Cu-S angle $\theta$ for Sr$_2$ScCuSO$_3$ without SOC.
The relaxation time $\tau=10^{-15}$ second was used for panel (a).}
\label{fig:param_angle}
\end{center}
\end{figure}

To quantify the change of the band structures, we investigated the relation between the S-Cu-S angle and PF$_{\mathrm{max}}$, $\Delta_{x^2-y^2}$ defined as the $d_{x^2-y^2}$ level relative to the $d_{xz/yz}$ level at the $\Gamma$ point, and the mass anisotropy $\gamma$ for Sr$_2$ScCuSO$_3$ without SOC.
Figure~\ref{fig:param_angle}(a)--(c) shows the plots, where both $\Delta_{x^2-y^2}$ and $\gamma$ increase by increasing the S-Cu-S angle. While PF$_{\mathrm{max}}$ is not so sensitive to the enhancement of the mass anisotropy, $\Delta_{x^2-y^2}\sim 0$ yields the pronounced PF$_{\mathrm{max}}$.

What is the microscopic origin of the change in the band structure?
For a simple limiting case: $\theta=180^\circ$, the Cu$Ch$ layer becomes a square lattice where Cu and $Ch$ atoms are alternatively aligned.
In this case, we can expect that the $d_{x^2-y^2}$ level will become higher than the $d_{xz/yz}$ level at the $\Gamma$ point for the following two reasons. First, the onsite energy of the $d_{x^2-y^2}$ orbital will be higher than that for the $d_{xz/yz}$ orbitals in that crystal field. Second, because of the two-dimensional conduction of the $d_{x^2-y^2}$ orbitals, the width of the $d_{x^2-y^2}$ band should be much larger than that for the $d_{xz/yz}$ bands, which is not necessarily the case when $\theta$ is much smaller than 180$^\circ$ because of relatively small hopping amplitudes between the $d_{x^2-y^2}$ and $p$ orbitals due to their different $c$ coordinates.
We also note that, for a small $\theta$, $d-d$ and $p-p$ hopping paths will also be activated.
In our tight-binding models, the onsite energy difference between the $d_{x^2-y^2}$ and $d_{xz/yz}$ orbitals is 0.04 eV for $\theta=113^\circ$ and 0.10 eV for $\theta=124^\circ$ in $Ch=$ S without SOC, so the variation in the on-site energy difference between the two cases is much smaller than the variance in $\Delta_{x^2-y^2}$, around 0.45 eV. Therefore, the main origin of the change in $\Delta_{x^2-y^2}$ is not the onsite energies of the $t_{2g}$ orbitals, but the other model parameters.
As a matter of fact, the $d_{x^2-y^2}$ band dispersion drastically changes by varying the bond angle $\theta$ as shown in Fig.~\ref{fig:band_angle}, which cannot be represented with a mere shift of the onsite energy.
We do not go into further detail because the required change in $\theta$ for the band degeneracy is prohibitively large: about 10$^\circ$ for $Ch=$ S, and $20^\circ$ for $Ch=$ Se.
While some atomic substitution, e.g., in the perovskite layer, might change the situation to some extent, we shall proceed in another way: looking for other candidate materials with Cu$Ch_4$ tetrahedra from Section.~\ref{sec:122}.

\subsubsection{Impact of the $d_{xz/yz}$ degeneracy on the power factor\label{sec:model_degen}}

Before proceeding, we take a brief look at the effect of the $d_{xz/yz}$ degeneracy at their band edge on the power factor, which is another important aspect of general $t_{2g}$ systems, while we have concentrated so far on the anisotropy (one-dimensionality) of the $d_{xz/yz}$ bands and the degeneracy between the $d_{xz/yz}$ and $d_{x^2-y^2}$ levels at the $\Gamma$ point.

To quantify this effect, we performed a simple model calculation for the following four situations:
(i) a single two-dimensional band dispersion:
\begin{equation}
\epsilon^{\mathrm{2D}} (k_x,k_y)=2t_1\cos k_x + 2t_2 \cos k_y, \label{eq:2D_1}
\end{equation}
(ii) a single three-dimensional band dispersion:
\begin{equation}
\epsilon^{\mathrm{3D}} (k_x,k_y,k_z)=2t_1\cos k_x + 2t_2 \cos k_y + 2t_2 \cos k_z, \label{eq:3D_1}
\end{equation}
(iii) degenerate two-dimensional band dispersions:
\begin{equation}
\begin{split}
\epsilon^{\mathrm{2D}}_1 (k_x,k_y)=2t_1\cos k_x + 2t_2 \cos k_y,\\
\epsilon^{\mathrm{2D}}_2 (k_x,k_y)=2t_2\cos k_x + 2t_1 \cos k_y, \label{eq:2D_2}
\end{split}
\end{equation}
and (iv) degenerate three-dimensional band dispersions,
\begin{equation}
\begin{split}
\epsilon^{\mathrm{3D}}_1 (k_x,k_y)=2t_1\cos k_x +2t_2 \cos k_y + 2t_2 \cos k_z,\\
\epsilon^{\mathrm{3D}}_2 (k_x,k_y)=2t_2 \cos k_x +2t_1 \cos k_y + 2t_2 \cos k_z. \label{eq:3D_2}
\end{split}
\end{equation}
Note that the term `degenerate' for the cases (iii) and (iv) is used to represent the degeneracy of the band edges of two dispersions at the $\Gamma$ point.
We set $t_1\geq t_2\geq 0$.
The mass anisotropy can be represented as $\gamma=t_1/t_2$.
We shall denote PF$_{\mathrm{max}}$ in each case (i)--(iv) as PF$^{\mathrm{2D}}_{\mathrm{max}}$, PF$^{\mathrm{3D}}_{\mathrm{max}}$, PF$^{\mathrm{2D, degen}}_{\mathrm{max}}$, and PF$^{\mathrm{3D, degen}}_{\mathrm{max}}$, respectively. The PF enhancement factor $\alpha$ is defined as follows:
\begin{equation}
\alpha_{\mathrm{2D}} \equiv \frac{\mathrm{PF}^{\mathrm{2D,degen}}_{\mathrm{max};xx}}{\mathrm{PF}^{\mathrm{2D}}_{\mathrm{max};xx}},\ \ 
\alpha_{\mathrm{3D}} \equiv \frac{\mathrm{PF}^{\mathrm{3D,degen}}_{\mathrm{max};xx}}{\mathrm{PF}^{\mathrm{3D}}_{\mathrm{max};xx}}, \label{eq:alpha}
\end{equation}
which quantify how much the degeneracy enhances PF$_{\mathrm{max}}$ along the $x$ direction.
All the results were obtained at the temperature $k_BT=0.1t_1$.

\begin{figure}
\begin{center}
\includegraphics[width=6.8 cm]{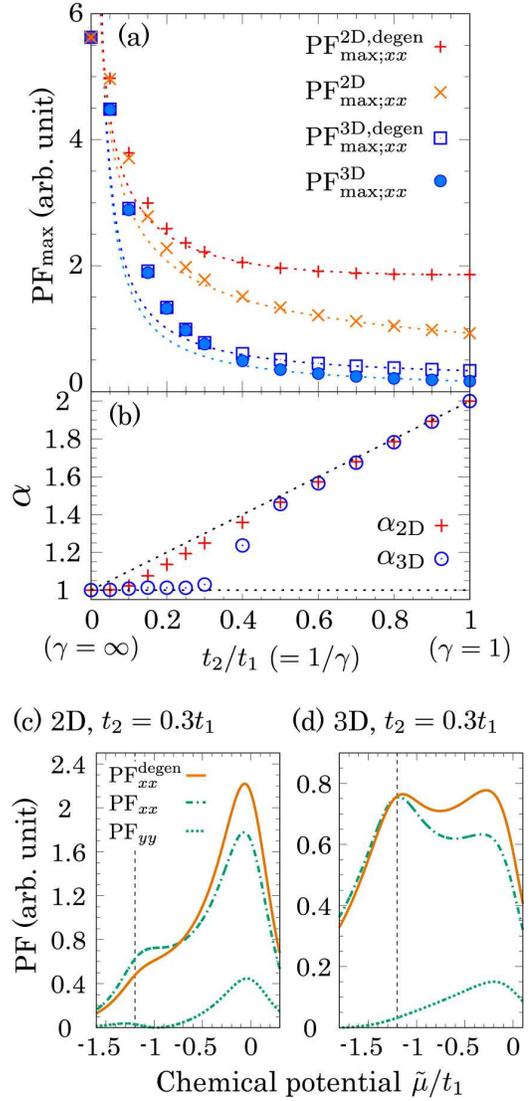}
\caption{(a) PF$_{\mathrm{max}}$ as a function of $t_2/t_1$. Dotted lines present the approximate relations Eqs.~(\ref{eq:apr1})--(\ref{eq:apr4}) for each case.
(b) The PF enhancement factor $\alpha$ defined in Eq.~(\ref{eq:alpha}) as a function of $t_2/t_1$. A dotted horizontal line shows $\alpha=1$, on which PF is not enhanced by degeneracy. The other dotted line shows the approximate relation Eq.~(\ref{eq:apr5}). (c)(d) Calculated PF as a function of the chemical potential relative to the valence-band top, $\tilde{\mu}$, normalized with $t_1$. Here, the results using $t_2=0.3t_1$ for the 2D and 3D band dispersions are shown in (c) and (d), respectively. The position of the van Hove singularity ($\tilde{\mu}/t_1=-1.2$) is shown with a broken line in each panel.}
\label{fig:model}
\end{center}
\end{figure}

PF$_{\mathrm{max}}$ and $\alpha$ with respect to the value of $t_2/t_1$ ($=\gamma^{-1}$) for each case are presented in Fig.~\ref{fig:model}(a) and (b), respectively. In these figures, we also show the following approximate relations, which are valid when $k_B T \ll t_1, t_2$ holds, by dotted lines:
\begin{eqnarray}
\mathrm{PF}^{\mathrm{2D}}_{\mathrm{max};xx} \simeq \sqrt{\gamma} \mathrm{PF}^{\mathrm{2D, iso}}_{\mathrm{max};xx},\label{eq:apr1}\\
\mathrm{PF}^{\mathrm{2D, degen}}_{\mathrm{max};xx} \simeq (\sqrt{\gamma}+\frac{1}{\sqrt{\gamma}}) \mathrm{PF}^{\mathrm{2D, iso}}_{\mathrm{max};xx},\label{eq:apr2}\\
\mathrm{PF}^{\mathrm{3D}}_{\mathrm{max};xx} \simeq \gamma \mathrm{PF}^{\mathrm{3D, iso}}_{\mathrm{max};xx},\label{eq:apr3}\\
\mathrm{PF}^{\mathrm{3D, degen}}_{\mathrm{max};xx} \simeq (\gamma + 1) \mathrm{PF}^{\mathrm{3D, iso}}_{\mathrm{max};xx},\label{eq:apr4}\\
\alpha_{\mathrm{2D}}\simeq \alpha_{\mathrm{3D}} \simeq 1+\frac{1}{\gamma},\label{eq:apr5}
\end{eqnarray}
where `iso' means the case of $t_2=t_1$. These relations are proven in APPENDIX.
Because the assumption required for these relations, $k_B T \ll t_1, t_2$, is satisfied for a large $t_2/t_1$ region in the plots shown in Fig.~\ref{fig:model}(a)--(b) since $k_BT = 0.1t_1$, these relations are valid in that region.

In Fig.~\ref{fig:model}(a), PF is actually increased by low dimensionality. This tendency is verified by comparing $\mathrm{PF}^{\mathrm{2D}}_{\mathrm{max};xx}$ and $\mathrm{PF}^{\mathrm{3D}}_{\mathrm{max};xx}$ in the whole region of $t_2/t_1$, or seeing a sharp increase of PF$_{\mathrm{max}}$ for $\gamma \to \infty$ ($t_2/t_1 \to 0$), which results in the one-dimensional band dispersion for all the four cases.
This increase is due to a large DOS near the edge for the low-dimensional band dispersion~\cite{Hicksone,Hicksone2,Dress,usuione}.

Next, we turn our attention to the effect of degeneracy. As is clearly seen in Fig.~\ref{fig:model}(b), the PF enhancement factor $\alpha$ is exactly two for $\gamma=1$ (isotropic band dispersion), but it decreases by increasing $\gamma$, and finally goes to unity for $\gamma=\infty$.
The reason for this behavior is explained as follows. For $\gamma=1$, two equivalent band dispersions are degenerate for the cases (iii) and (iv), which simply doubles the transport coefficients ${\bf K}_{\nu}$ compared with those for the cases (i) and (ii). Thus, PF $\propto K_1^2 / K_0$ is also doubled, which means $\alpha=2$. On the other hand, for the cases (iii) and (iv) with $\gamma=0$, there are one-dimensional band dispersions with respect to the two different directions.
In spite of the degeneracy at the band edge, each band dispersion can contribute to the transport only along the one direction.
For example, in the case (iii), PF$_{xx}$ is determined solely by $\epsilon^{\mathrm{2D}}_1 (k_x,k_y)=2t_1\cos k_x$, and $\epsilon^{\mathrm{2D}}_2 (k_x,k_y)=2t_1\cos k_y$ makes no contribution for it. Therefore, the degeneracy plays no role for PF enhancement: $\alpha=1$.

In Fig.~\ref{fig:model}(b), we can also find that $\alpha_{\mathrm{3D}}$ goes down more rapidly than $\alpha_{\mathrm{2D}}$.
To investigate this difference, PF as a function of the chemical potential relative to the valence-band top, $\tilde{\mu}$, for the 2D and 3D cases with $t_2=0.3t_1$ is presented in Fig.~\ref{fig:model}(c) and (d), respectively.
For the 2D band dispersion, both PF$_{xx}$ and PF$_{yy}$ in the case (i) have a maximum near the band edge ($\tilde{\mu}=0$), and then PF$_{xx}^{\mathrm{degen}}$ in the case (iii) also has a peak there, where the peak value is approximately represented as a summation of PF$_{xx}$ and PF$_{yy}$ (see APPENDIX).
On the other hand, for the 3D band dispersion, while PF$_{yy}$ in the case (ii) becomes maximum near the band edge, PF$_{xx}$ has a peak near the van Hove singularity at $\tilde{\mu}/t_1 = -1.2$. This is the origin of the small $\alpha_{\mathrm{3D}}$, in other words, weak enhancement of the PF$_{xx}^{\mathrm{degen}}$ peak value compared with PF$_{xx}$.
The difference originates from a sharp increase of DOS at the edge of the two-dimensional band dispersion, which is absent at the edge of the three-dimensional band dispersion.
We note that (i) when one focuses on PF near the band edge, the enhancement is observed to some extent, and (ii) the existence of the van Hove singularity is not a universal issue for general systems.
In addition, one often has to concentrate on the PF near the band edge in order to get a $ZT$ peak because a deep chemical potential usually yields an increase in the electronic thermal conductivity, which decreases $ZT$.
Because of these reasons, the difference between 2D and 3D shown in Fig.~\ref{fig:model}(b) is not so general.

Here, we come back to our first-principles calculation, which yields $\gamma=6.2$ and 9.0 for $Ch=$ S and Se in Sr$_2$ScCu$Ch$O$_3$, respectively.
Because of such a large anisotropy and the observation made in this subsection, we can conclude that the degeneracy of the $d_{xz/yz}$ band edges is not so effective to enhance the power factor, while the degeneracy between the $d_{xz/yz}$ and $d_{x^2-y^2}$ band edges can drastically increase the power factor. In general, degeneracy of the band dispersions with strong anisotropy with respect to the orthogonal directions cannot increase the power factor so much.

\subsection{$\beta$-BaCu$_2$S$_2$\label{sec:122}}

\begin{figure}
\begin{center}
\includegraphics[width=8.2 cm]{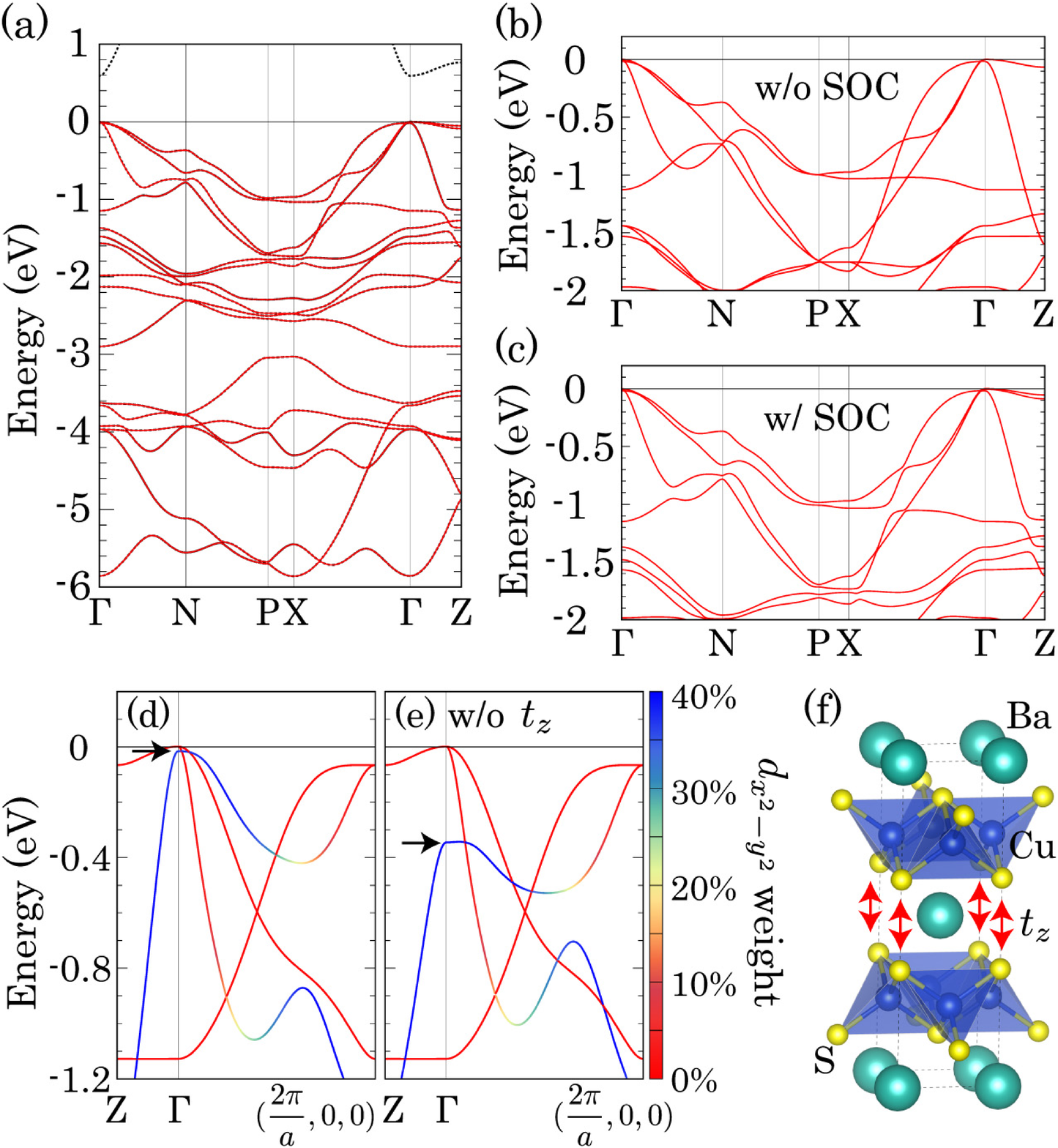}
\caption{(a) Band structure of $\beta$-BaCu$_2$S$_2$ calculated with SOC. Black broken and red solid lines represent the band structures obtained with the first-principles calculation and the tight-binding model of the Wannier functions, respectively. Blow-up views of the first-principles band structures near the valence-band top calculated (b) without and (c) with SOC. (d) Tight-binding band structure calculated without SOC, where the $d_{x^2-y^2}$ orbital weight is shown by color. An arrow denotes the edge of the $d_{x^2-y^2}$ band. The band dispersion along the Z-$\Gamma$-($k_x$, $k_y$, $k_z$)=($2\pi/a$, 0, 0) line is shown. (e) The same as panel (d) but the $t_z$ hopping is hypothetically removed from our tight-binding model. (f) Definition of the $t_z$ hopping between the S-$p_z$ orbitals.}
\label{fig:band_BaCu2S2}
\end{center}
\end{figure}

\begin{figure}
\begin{center}
\includegraphics[width=8 cm]{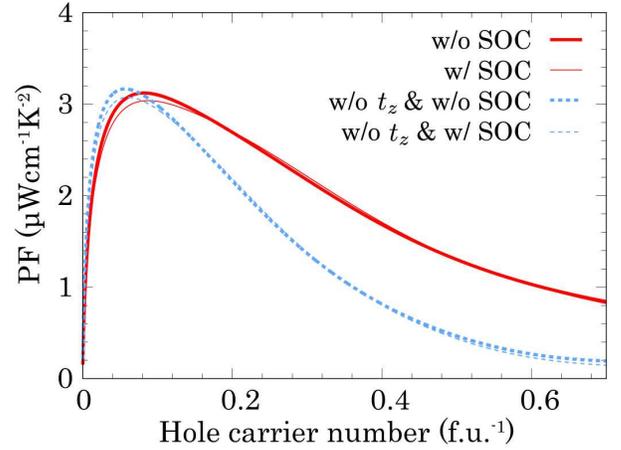}
\caption{\label{fig:PF_BaCu2S2} Calculated PF of $\beta$-BaCu$_2$S$_2$ for several conditions: a regular calculation and one without $t_z$, with and without SOC. These results are shown using red thick solid (regular calculation without SOC), red thin solid (regular calculation with SOC), blue thick broken (without $t_z$ and without SOC), and blue thin broken lines (without $t_z$ and with SOC), respectively. The relaxation time $\tau=10^{-15}$ second was used.}
\end{center}
\end{figure}

From now on, we investigate the electronic structure of other materials with Cu$Ch_4$ tetrahedra.
The first target is $\beta$-BaCu$_2$S$_2$ with separated CuS layers similarly to Sr$_2$ScCuSO$_3$.
Figure~\ref{fig:band_BaCu2S2}(a) presents the first-principles (black broken lines) and tight-binding (red solid lines) band structures of $\beta$-BaCu$_2$S$_2$ calculated with SOC. Blow-up views of the first-principles band structures near the valence-band top calculated without and with SOC are shown in Fig.~\ref{fig:band_BaCu2S2}(b) and (c), respectively.
In spite of the similarity of the crystal structure to that of Sr$_2$ScCuSO$_3$, a remarkable difference in the band structure is a nearly degenerate $d_{x^2-y^2}$ band edge at the $\Gamma$ point in $\beta$-BaCu$_2$S$_2$.
This degeneracy can be seen more easily in Fig.~\ref{fig:band_BaCu2S2}(d), where the tight-binding band structure calculated without SOC is shown with the $d_{x^2-y^2}$ orbital weight by color. An arrow in the figure denotes the $d_{x^2-y^2}$ band edge at the $\Gamma$ point.
To investigate the effect of the band degeneracy on the power factor, in Fig.~\ref{fig:band_BaCu2S2}(e), we also present the tight-binding band structure without an inclusion of the $t_z$ hopping between the S-$p_z$ orbitals, which is defined in Fig.~\ref{fig:band_BaCu2S2}(f).
As is clearly seen in Fig.~\ref{fig:band_BaCu2S2}(e), the $d_{x^2-y^2}$ band edge is lowered by the hypothetical removal of the $t_z$ hopping, while the $d_{xz/yz}$ band structures are not changed so much. By comparing the power factors obtained from these two conditions, we can evaluate the effect of the $d_{x^2-y^2}$ band degeneracy on the thermoelectric performance. We note that the Cu-$d_{x^2-y^2}$ orbitals strongly hybridize with the S-$p_z$ orbitals, as seen in a sharp band dispersion of the $d_{x^2-y^2}$ band along the $\Gamma$-Z line, while the hybridization of the S-$p_z$ and Cu-$d_{xz/yz}$ orbitals is prohibited at the $\Gamma$ point as we have seen for Sr$_2$ScCuSO$_3$.
Such an inter-layer hopping is almost absent in Sr$_2$ScCuSO$_3$.

Figure~\ref{fig:PF_BaCu2S2} shows the calculated PF for several conditions: a regular calculation and one without $t_z$, with and without SOC.
It is natural that the SOC plays a negligible role on the power factor.
However, against our expectation, PF$_{\mathrm{max}}$ is not enhanced by the degeneracy of the band edge.
A possible reason for this is the sizable anisotropy, i.e. the quasi-one-dimensionality, of all the $t_{2g}$ orbitals.
For example, the $d_{xz/yz}$ bands are very heavy along the $z$ direction, and the $d_{x^2-y^2}$ band is much heavier along the $x$ and $y$ directions than the $z$ direction. Therefore, the power factor along the $x$ direction is not as improved by the $d_{x^2-y^2}$ band with a sharp dispersion along the $z$ direction, as discussed in Sec.~\ref{sec:model_degen}.
Note that this characteristics is not inherent to the compounds with the same or similar structures.
As a matter of fact, the $d_{xz/yz}$ bands have a sizable dispersion along the $z$ direction for other materials such as Mg$_3$Sb$_2$, and in that case, the degeneracy becomes a good measure for evaluating its power factor as shown in Ref.~[\onlinecite{122_degen}].
The higher value of the calculated PF of $\beta$-BaCu$_2$S$_2$ than that of Sr$_2$ScCuSO$_3$ might be due to a large difference of the band width: the band structures for the former compound exhibit a much sharper dispersion as presented in Figs.~\ref{fig:band_pero}(c) and \ref{fig:band_BaCu2S2}(d).

\subsection{BiCuSeO\label{sec:BiCuSeO}}

BiCuSeO is one of the most famous thermoelectric materials with Cu$Ch$ layers, but its electronic structure is rather exceptional.
In previous studies, it is shown that the Bi atoms play a crucial role not only in reducing its lattice thermal conductivity~\cite{BiCuSeO_phonon}
but also in changing the shape of its band structure drastically~\cite{BiCuSeO_band_deform1,BiCuSeO_band_deform2,BiCuSeO_band_deform3,BiCuSeO_band_deform4,BiCuSeO_boltz}. 

\begin{figure}
\begin{center}
\includegraphics[width=7.5 cm]{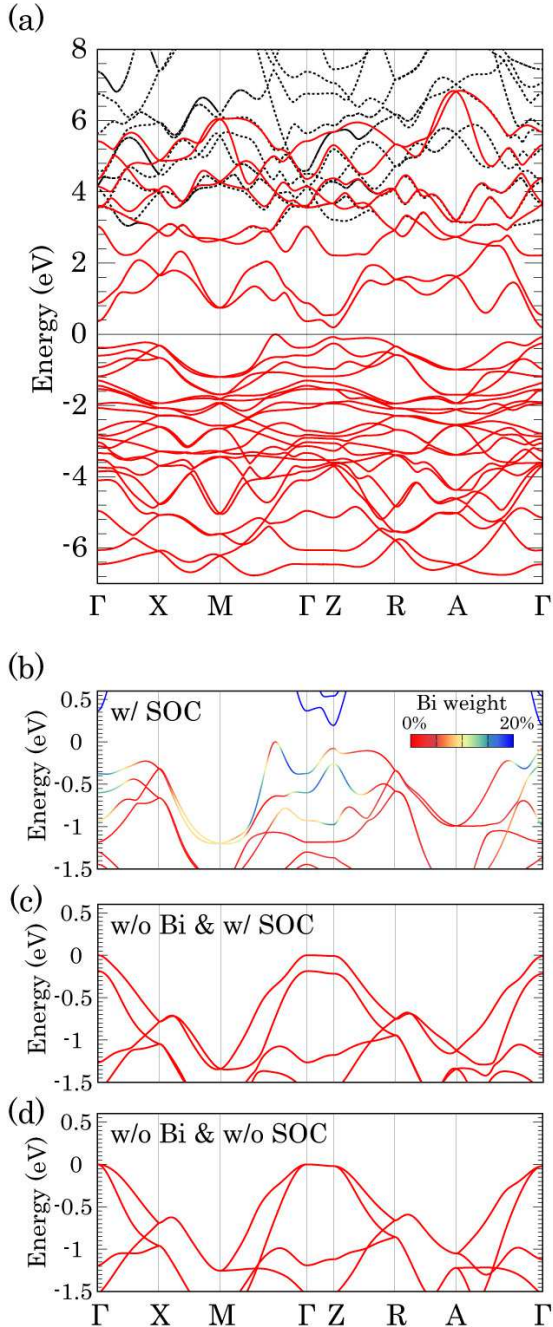}
\caption{(a) Band structures of BiCuSeO calculated with SOC, where the black broken and red solid lines present the first-principles band structure and that obtained by using the tight-binding model consisting of the Wannier functions, respectively. (b) Blow-up view of the band structure calculated using the tight-binding model, where the Bi orbital weight is shown by color.
(c) The same band structure calculated with a hypothetical removal of the Bi orbitals contribution. (d) The same plot as (c) but calculated without SOC.}
\label{fig:band_BiCuSeO}
\end{center}
\end{figure}

Figure~\ref{fig:band_BiCuSeO}(a) shows band structures calculated with SOC, where the black broken and red solid lines present those obtained with the first-principles and tight-binding model calculations, respectively.
The blow-up view of the band structure calculated using the tight-binding model is presented in Fig.~\ref{fig:band_BiCuSeO}(b), where the Bi orbital weight is shown by color.
As is clearly seen, the Bi states are strongly hybridized with the valence-band edge near the $\Gamma$ point. As a result, a peculiar shape of the band structure is realized here.
If one hypothetically eliminates the Bi orbitals contribution from the band structure, the valence-band structures shown in Fig.~\ref{fig:band_BiCuSeO}(c)--(d) are very similar to that for Sr$_2$ScCu$Ch$O$_3$.

\begin{figure}
\begin{center}
\includegraphics[width=8 cm]{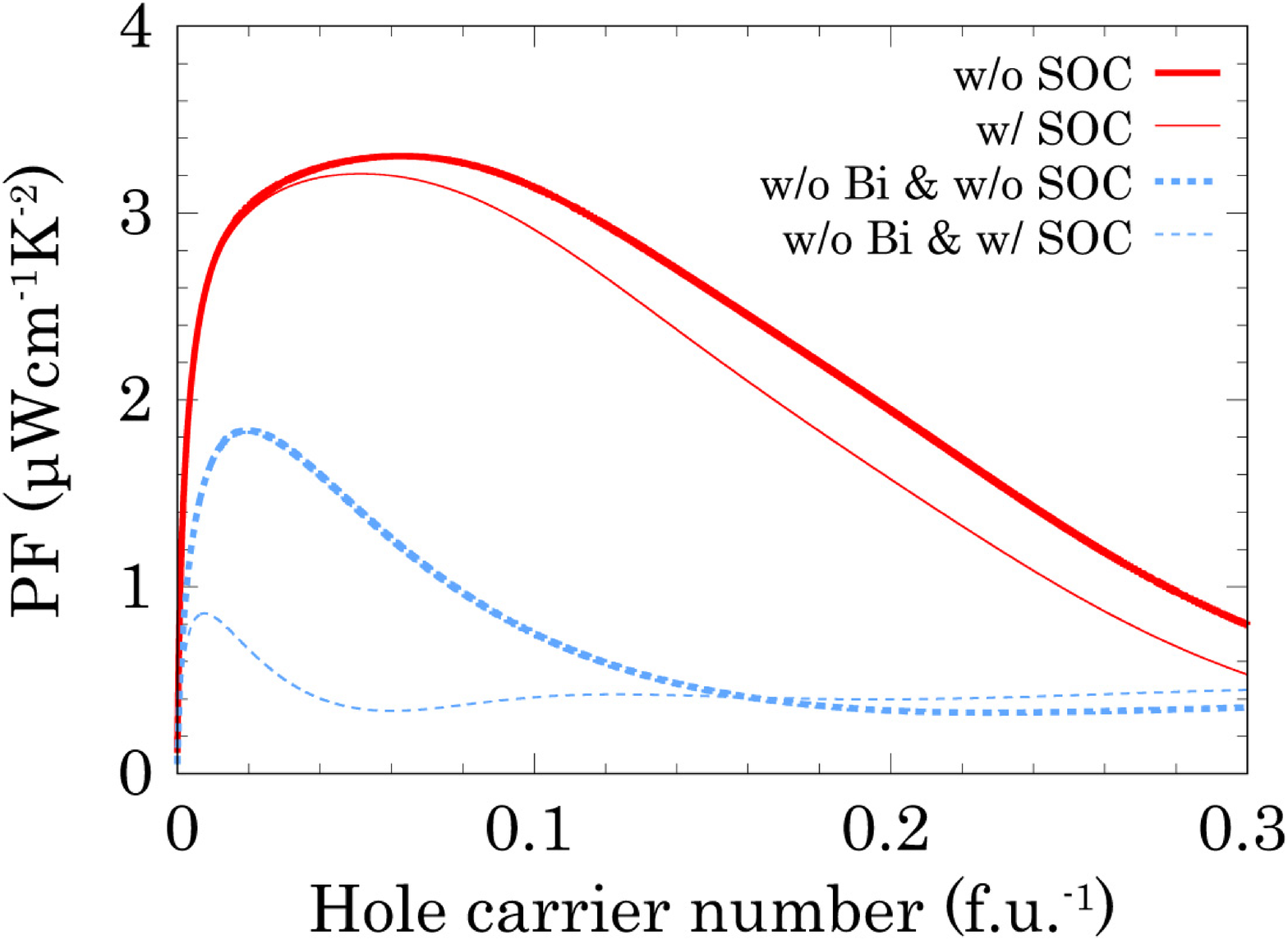}
\caption{\label{fig:PF_BiCuSeO} Calculated PF of BiCuSeO for several conditions: a regular calculation and one without the contribution of the Bi orbitals (denoted as `w/o Bi'), with and without SOC. These results are shown using red thick solid (regular calculation without SOC), red thin solid (regular calculation with SOC), blue thick broken (without Bi and without SOC), and blue thin broken lines (without Bi and with SOC), respectively.
The relaxation time $\tau=10^{-15}$ second was used.}
\end{center}
\end{figure}

Calculated PF is shown in Fig.~\ref{fig:PF_BiCuSeO}, where we also show PF with a hypothetical removal of the Bi orbitals contribution from our tight-binding model. The calculation results with and without SOC are plotted. 
Here, we neglected the conduction bands in transport calculations because GGA is known to underestimate the band gap.
Because the experimental direct band gap is about 0.8 eV~\cite{BiCuSeO_band_deform1}, we can safely neglect the conduction bands in evaluating PF at 300 K~\cite{footnote_BiCuSeO}.
In Fig.~\ref{fig:PF_BiCuSeO}, we can verify that the Bi hybridization drastically increases PF$_{\mathrm{max}}$ by a factor of three when calculation includes SOC. As seen in Fig.~\ref{fig:band_BiCuSeO}(b), the valence-band deformation can yield a large valley degeneracy, the Fermi surface for which was shown in Ref.~[\onlinecite{BiCuSeO_band_deform4}].
This is the origin of the large enhancement of PF while the band dispersion along the $\Gamma$-Z line suggests a rather isotropic electronic structure (i.e.~no longer flat along the $\Gamma$-Z line), as pointed out by previous studies (e.g.~Refs.~[\onlinecite{BiCuSeO_band_deform2, BiCuSeO_band_deform4}]).
On the other hand, the power factor without the Bi hybridization is rather similar to those obtained for Sr$_2$ScCu$Ch$O$_3$.
In fact, the mass anisotropy $\gamma$ for the band structure shown in Fig.~\ref{fig:band_BiCuSeO}(d) is around 4.2, which is not a large value.
Therefore, Bi hybridization with the valence-band top is crucial in BiCuSeO to enhance its PF.

\subsection{Zincblende and wurtzite Cu$Ch$\label{sec:ZBWZ}}
 
We have focused so far on the two-dimensional Cu$Ch$ layered structures, but how about the three-dimensional network of copper and chalcogen atoms? 
Zincblende and wurtzite are the most fundamental crystal structures consisting of tetrahedra.
Because of such an importance, we investigated zincblende and wurtzite Cu$Ch$, even though they are hypothetical materials. 
While zincblende and wurtzite Zn$Ch$ exist as insulators, we adopted Cu$Ch$ as our target to make a fair comparison between the electronic structure and those in other materials with Cu$Ch_4$ tetrahedra investigated in this paper.

\begin{figure}
\begin{center}
\includegraphics[width=8.5 cm]{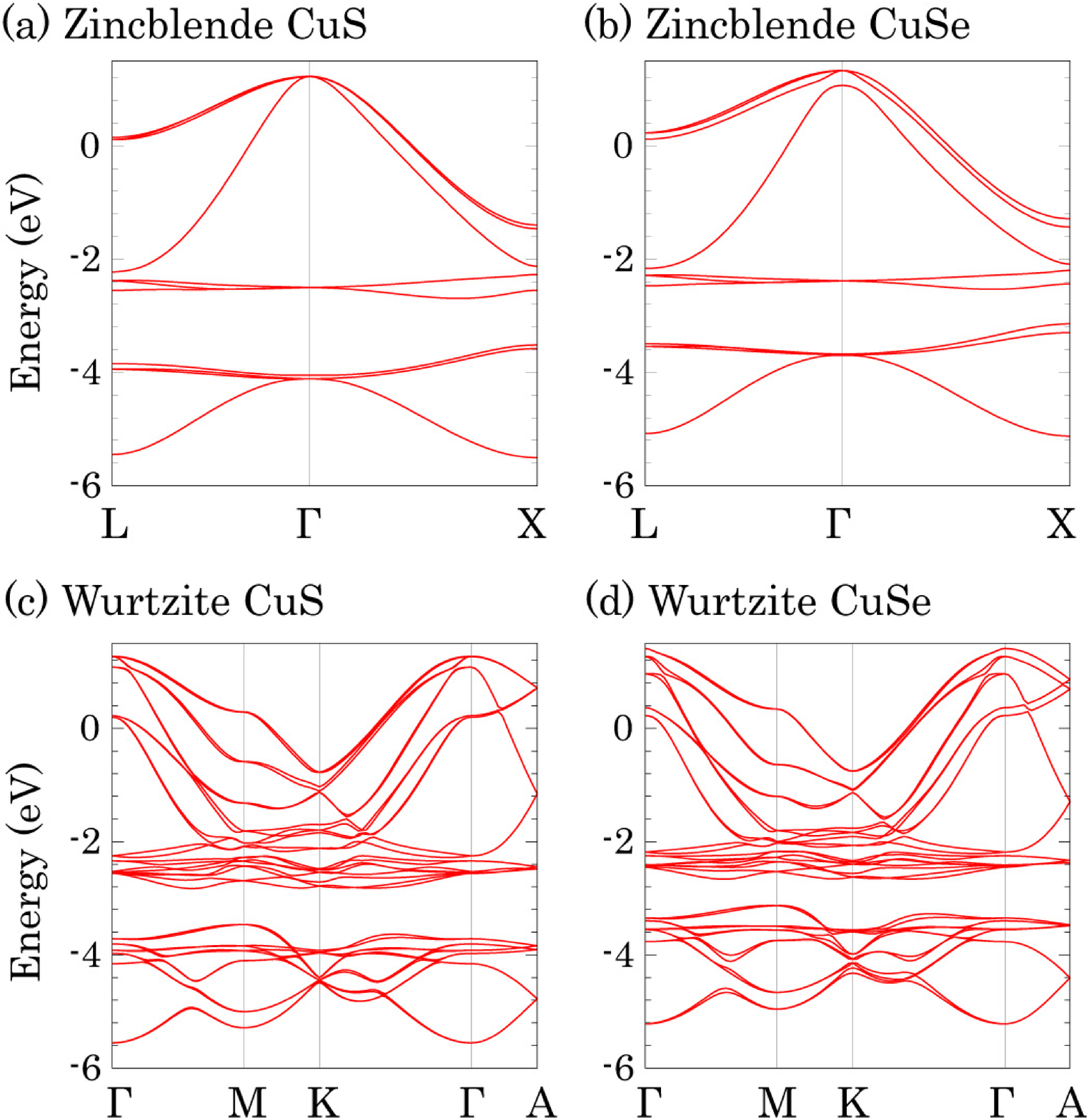}
\caption{First-principles band structures of (a) zincblende CuS, (b) zincblende CuSe, (C) wurtzite CuS, and (d) wurtzite CuSe, respectively, calculated with SOC.}
\label{fig:band_ZBWZ}
\end{center}
\end{figure} 

Figure~\ref{fig:band_ZBWZ}(a)--(d) present the first-principles band structures of zincblende CuS, zincblende CuSe, wurtzite CuS, and wurtzite CuSe, respectively, calculated with SOC.
All of these compounds have the valence-band top at the $\Gamma$ point that consists of the Cu-$t_{2g}$ orbitals hybridized with the $Ch$-$p$ orbitals.
We note that, in these systems, the onsite energies of the Cu-$t_{2g}$ and $Ch$-$p$ orbitals are very close. Whereas it is not very clear whether the band structure near the valence-band top should be called Cu-$t_{2g}$ or the $Ch$-$p$ bands, we call them Cu-$t_{2g}$ in a similar manner to other materials in this study.
Whereas these materials have three-dimensional crystal structures, the band dispersions still have anisotropy to some extent because of the orbital anisotropy. For example, in Fig.~\ref{fig:band_ZBWZ}(a), there are two degenerate heavy bands and one light band along the $\bm{k}$-path between the L $(k_x,k_y,k_z)=(\pi/a, \pi/a, -\pi/a)$ and $\Gamma$ points, which corresponds to the quasi-one-dimensionality.
Note that the one-dimensionality is rather weak, as expected: $\gamma=$ 2.2 and 3.5 for $Ch=$ S and Se with the zincblende structure, respectively, where SOC is not included in this evaluation.
The high symmetry of the zincblende structure results in a complete degeneracy of the $t_{2g}$ band edges unless SOC is switched on.
The wurtzite structure does not exhibit such a degeneracy because of a lack of the corresponding symmetry, and as a result, the $d_{x^2-y^2}$ band lies below the valence-band edge, as shown in Fig.~\ref{fig:band_ZBWZ}(c)--(d).
The effect of SOC can be inferred from the difference between CuS and CuSe, which acts in the same manner as in Sr$_2$ScCu$Ch$O$_3$.

\begin{figure}
\begin{center}
\includegraphics[width=8.6 cm]{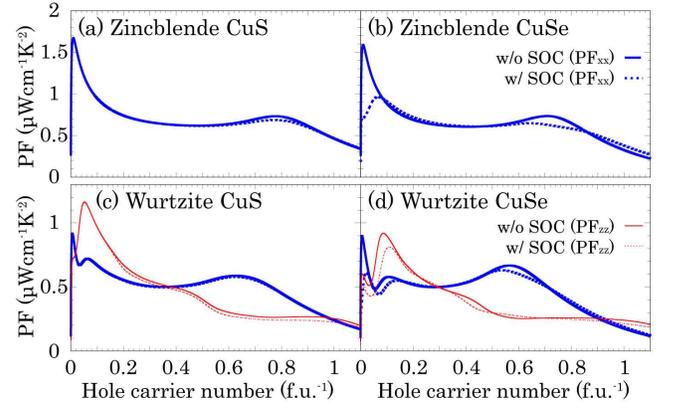}
\caption{Calculated PF with respect to the hole carrier number of (a) zincblende CuS, (b) zincblende CuSe, (c) wurtzite CuS, and (d) wurtzite CuSe, respectively. Calculation results obtained with and without SOC are shown using thin broken and solid lines, respectively. For (c)--(d), thick blue and thin red lines correspond to PF$_{xx}$ and PF$_{zz}$, respectively, while they are equivalent for the zincblende structure.
The relaxation time $\tau=10^{-15}$ second was used for all the panels. The zero hole carrier number corresponds to the insulating state where all the bands shown in Fig.~\ref{fig:band_ZBWZ} are filled.}
\label{fig:PF_ZBWZ}
\end{center}
\end{figure}

The calculated PF with respect to the hole carrier number is shown in Fig.~\ref{fig:PF_ZBWZ}.
The zero hole carrier number corresponds to the insulating state where all the bands shown in Fig.~\ref{fig:band_ZBWZ} are filled.
For zincblende Cu$Ch$, PF$_{\mathrm{max}}$ reaches comparable values with layered compounds we have investigated in this paper, although the anisotropy is weaker in the sense that electrons are mobile along the $z$ direction here.
One of the possible origins is the three-fold degeneracy of the quasi-one-dimensional $t_{2g}$ bands.
While the degeneracy of the quasi-one-dimensional band dispersions only weakly enhances PF, as seen in Sec.~\ref{sec:model_degen},
 the PF enhancement factor $\alpha$ for weakly anisotropic band dispersions, which is roughly estimated as $1+2\gamma^{-1}$, amounts to 1.9 and 1.6 for $Ch=$ S and Se, respectively.
Therefore, in spite of the weak anisotropy originating from the three-dimensional network, the three-fold degeneracy thanks to the high crystal symmetry is rather effective to increase PF of zincblende Ch$Ch$.
For wurtzite Cu$Ch$, PF$_{\mathrm{max}}$ are much lower.
This might be due to the lack of the $t_{2g}$ degeneracy together with a sizable transfer along the $c$-axis, which deteriorates the anisotropy of the electronic structure.
For both crystal structures, CuS is almost unaffected by SOC while PF$_{\mathrm{max}}$ for CuSe exhibits a sizable reduction by SOC.
This tendency is exists also in the materials we have investigated before in this paper because SOC acts on the band structure in the same manner.
In summary, we can say that the valence-band structures can be understood in the same manner as those in the layered compounds: degeneracy and anisotropy of the Cu-$t_{2g}$ bands near the $\Gamma$ point determine PF while SOC deteriorates them for relatively heavy atoms.

\subsection{Cu$_{12}$Sb$_4$S$_{13}$ (tetrahedrite)\label{sec:tetra}}
 
\begin{figure*}
\begin{center}
\includegraphics[width=17 cm]{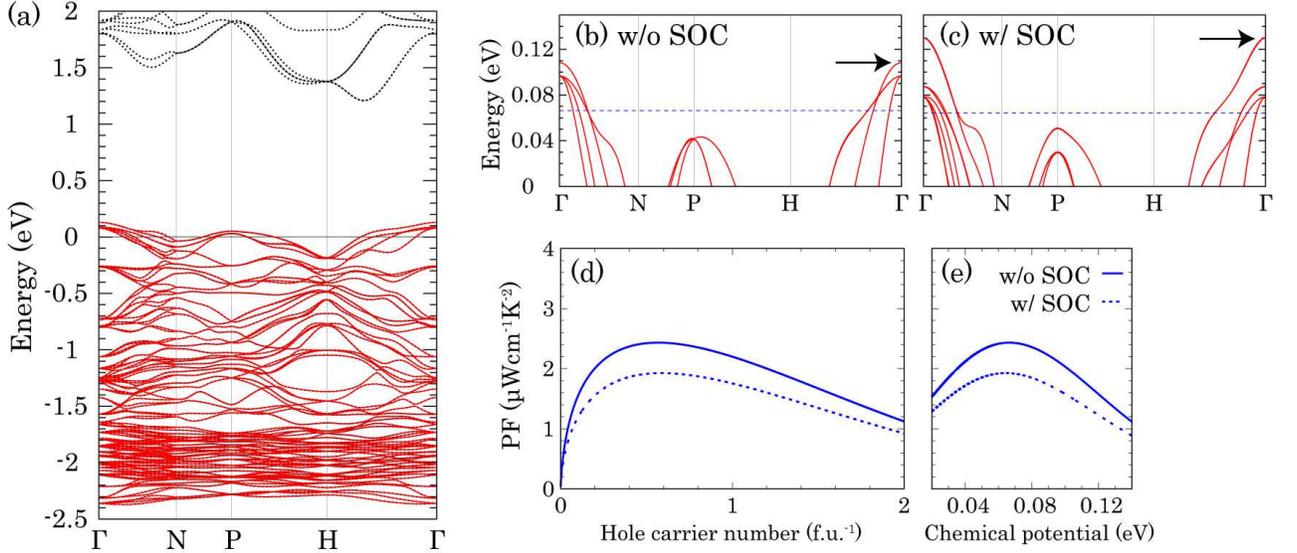}
\caption{(a) Band structures of Cu$_{12}$Sb$_4$S$_{13}$ obtained with SOC, where the black broken and red solid lines present the first-principles calculations and tight-binding models consisting of the Wannier functions, respectively. Blow-up view of the first-principles band structures calculated (b) without and (c) with SOC.
The calculated PF (d) with respect to the hole carrier number and (e) with respect to the chemical potential. In (d)--(e), calculation results obtained with and without SOC are shown using thin broken and thick solid lines, respectively.
The zero hole carrier number in (d) corresponds to the insulating state where all the bands shown with the red solid lines in (a) are filled.
Chemical potentials that maximize PF in (e) are shown in (b)--(c) with blue broken lines.
The relaxation time $\tau=10^{-15}$ second was used for (d) --(e).}
\label{fig:tetra}
\end{center}
\end{figure*}

Tetrahedrite Cu$_{12}$Sb$_4$S$_{13}$ has a complex crystal structure as shown in Fig.~\ref{fig:crystal}(f), which contains CuS$_4$ tetrahedra together with CuS$_3$ triangles.
Here, we considered the crystal structure with a space group $I\bar{4}3m$~\cite{tetrahed_strct}.
While the lower symmetry phase was reported in low temperature below 88 K by experiments~\cite{tetrahed_lowsym}, we concentrate on the cubic phase to investigate its thermoelectric performance.
Figure~\ref{fig:tetra}(a) presents the first-principles (black broken lines) and tight-binding (red solid lines) band structures for Cu$_{12}$Sb$_4$S$_{13}$. We note that the stoichiometric Cu$_{12}$Sb$_4$S$_{13}$ is metallic.
Blow-up views of the first-principles band structures are shown in Fig.~\ref{fig:tetra}(b)--(c) without and with SOC, respectively.
By looking into the $\bm{k}$-path between the H $(k_x,k_y,k_z)=(2\pi/a, 0, 0)$
and $\Gamma$ points in Fig.~\ref{fig:tetra}(b), we can see that there are three quasi-one-dimensional band dispersions and one isotropic band dispersion, the latter of which lies in the valence-band top as indicated by an arrow.
Note that there is a two-fold degeneracy for the band dispersion with the heavy effective mass of the quasi-one-dimensional band dispersions along the H-$\Gamma$ line, in a similar manner to the band dispersions of the zincblende Cu$Ch$ along the L-$\Gamma$ line.
SOC makes the relative energy level of the isotropic band compared with the quasi-one-dimensional bands higher by around 30 meV.

The calculated PF is shown in Fig.~\ref{fig:tetra}(d)--(e), with respect to the hole carrier number and the chemical potential, respectively.
The zero hole carrier number in Fig.~\ref{fig:tetra}(d) corresponds to the insulating state where all the bands shown with the red solid lines in Fig.~\ref{fig:tetra}(a) are filled.
The obtained PF is not small in spite of the weak anisotropy compared with layered structures in the sense that electrons are mobile along the $z$ direction. A good hint to understand this behavior is
a chemical potential corresponding to PF$_{\mathrm{max}}$, around 60 or 70 meV, which is shown with blue broken lines in Fig.~\ref{fig:tetra}(b)--(c).
This is a bit far from the band edge, and so a relatively small splitting between the quasi-one-dimensional and isotropic band dispersions at the $\Gamma$ point might not cause a sizable reduction in PF.
In addition, we can also notice that another valley around the P point possibly contributes to PF~\cite{tetrahed_review}.
While the basic characteristics of the band structure are similar to that for other materials with Cu$Ch_4$ tetrahedra,
this interesting multiple degeneracy for these complex band structures seems to contribute to enhance PF in this material.

\subsection{Other materials with Cu$Ch_4$ tetrahedra\label{sec:others}}

One of the other important examples of materials with Cu$Ch_4$ tetrahedra is Cu$_2Ch$, which is known to exhibit a high $ZT$~\cite{Cu2Ch_expt1,Cu2Ch_expt2,Cu2Se_expt} such as 1.5 at 1000 K for Cu$_{2-x}$Se~\cite{Cu2Se_expt}.
While we do not investigate this class of material due to its theoretical complexity including a large sensitivity of its band structure to the calculation methods~\cite{Cu2Ch_thr}, we point out here that its valence-band structure near the band edge (e.g. see Ref.~[\onlinecite{Cu2Ch_thr}]) looks similar to that of the hypothetical zincblende Cu$Ch$ shown in our paper because of the similarity in the crystal structure.

\section{Conclusion\label{sec:sum}}

In this study, we have performed a comparative study on the thermoelectric performance of materials with Cu$Ch_4$ tetrahedra.
We have found that their electronic band structure is characterized as a degenerate low-dimensional electronic structure of the Cu-$t_{2g}$ states.
In particular, for materials with separate Cu$Ch$ layers, the Cu-$t_{2g}$ bands are further classified as quasi-one-dimensional Cu-$d_{xz/yz}$ and two-dimensional Cu-$d_{x^2-y^2}$ states, the degeneracy of which can be crucial to improve their thermoelectric performance. However, it is sometimes not effective, as in $\beta$-BaCu$_2$S$_2$, where the Cu-$d_{x^2-y^2}$ band with a large out-of-plane group velocity and a small in-plane group velocity does not enhance the in-plane power factor.
By looking into the important hopping paths for determining the low dimensionality shown in Figs.~\ref{fig:hop}(b)--(c) and \ref{fig:hop_change}(b),
we can say that the Cu$Ch$ layer can be regarded as a kind of square lattice for both the $d$-$p$ and $p$-$p$ networks.

One of the interesting characteristics of Cu$Ch_4$ tetrahedra is the chalcogen orbitals extended in the void region, which enables a relatively distant transfer of electrons.
Such an extended chalcogen orbital sometimes forms hidden chains in the layered compounds, which strongly enhances the one-dimensionality of the $d_{xz/yz}$ band dispersion, and sometimes induces a strong inter-layer coupling, which alters the valence-band structure of BiCuSeO and $\beta$-BaCu$_2$S$_2$ and then largely improves the thermoelectric performance for BiCuSeO. 
In other words, the important aspects of the band structures of materials with Cu$Ch_4$ tetrahedra often originate from the existence of the extended chalcogen orbitals, which sometimes makes the blocking layer rather crucial for thermoelectric performance. In fact, while the $t_{2g}$ band degeneracy looks important for enhancing the power factor of Sr$_2$ScCu$Ch$O$_3$, this is not the case for $\beta$-BaCu$_2$S$_2$. In addition, the electronic structure of the CuSe layer is largely altered by the BiO layer in BiCuSeO. Controlling the blocking layer can bring a large modification of the electronic structure in materials with Cu$Ch$ layers.

The large extension of the chalcogen orbitals is partially due to the anisotropic Cu-$Ch$ bond geometry of a tetrahedron.
This situation reminds us of the concept of a lone-pair electron~\cite{lone_pair1,lone_pair2,lone_pair3}, which extends into the void space partially by the anisotropic environment.
While a crucial role of lone-pair electrons on structural instability in realizing low thermal conductivity has been pointed out for several (thermoelectric) materials~\cite{lone_pair1,lone_pair2,lone_pair3}, it is interesting that the extended electronic states (but not the lone-pair electrons) can also enhance the power factor by altering the electronic band structure.
In BiCuSeO, the origin of the alternation of the valence-band structure might be interpreted as an interplay of the lone-pair electrons of the Bi atoms and the extended Se orbitals.

Another characteristic of the tetrahedron is the large band splitting by SOC, which might be because of the lack of inversion symmetry of a tetrahedron. While the power factor is not as degraded by SOC for Se at room temperature, a much higher temperature will be desirable to make the SOC band splitting deteriorate the power factor less by large temperature broadening for materials with heavier atoms such as Te.

Several observations made in this study will be important, not only for designing new thermoelectric materials with Cu$Ch_4$ tetrahedra
but also for investigating the interesting and complicated roles of extended electronic states on the transport properties of electrons and phonons.

\acknowledgments
We appreciate fruitful discussion with Hiraku Ogino, Koichiro Suekuni, Kunihiro Kihou, and Chul-Ho Lee. 
This study was supported by JSPS KAKENHI (Grant Nos.~JP17H05481 and JP17K14108) and JST CREST (Grant No.~JPMJCR16Q6), Japan.

\section*{APPENDIX: Powerfactor for degenerate anisotropic bands}
In this section, we assume that $k_BT/t_1$ is always fixed.
First, we consider the single-band problem with anisotropy. If the band dispersion is given as Eq.~(\ref{eq:2D_1}),
the transport coefficients along the $x$ direction read
\begin{equation}
\scalebox{0.9}{$\displaystyle
K^{\mathrm{2D}} _{xx; \nu} [\mu]= \tau \int \mathrm{d}\bm{k}\ (4t_1^2\sin^2 k_x) g_{\nu}(2t_1\cos k_x + 2t_2 \cos k_y-\mu) ,
$}
\end{equation}
where
\begin{equation}
g_{\nu}(E)\equiv \left[-\frac{\partial f_0}{\partial E} (\frac{E}{k_B T})\right]E^\nu .
\end{equation}
As described in the main text, the anisotropy $\gamma$ is defined as $\gamma=t_1/t_2$.
If the temperature is sufficiently smaller than $t_1$ and $t_2$ (i.e. $k_B T \ll t_1,t_2$), the transport coefficients are approximately related to those for the isotropic case ($t_2=t_1$), $K^{\mathrm{2D, iso}}_{xx; \nu}$, as follows:
\begin{equation}
K^{\mathrm{2D}} _{xx;\nu}[\mu] \simeq \sqrt{\gamma}K^{\mathrm{2D, iso}}_{xx;\nu}[\mu+2(t_2-t_1)],
\end{equation}
by using $\cos k_y \simeq 1-k_y^2/2$ near the $\Gamma$ point and a transformation $\tilde{k}_y = k_y/ \sqrt{\gamma}$.
Thus, PF$_{\mathrm{max}}$, which is the maximum value of the power factor with respect to the chemical potential, satisfies
\begin{equation}
\mathrm{PF}^{\mathrm{2D}} _{\mathrm{max};xx} \simeq \sqrt{\gamma} \mathrm{PF}^{\mathrm{2D, iso}}_{\mathrm{max};xx}.
\end{equation}
In the same manner, we can obtain
\begin{equation}
\mathrm{PF}^{\mathrm{2D}} _{\mathrm{max};yy} \simeq \frac{1}{\sqrt{\gamma}} \mathrm{PF}^{\mathrm{2D, iso}}_{\mathrm{max};yy},
\end{equation}
by using $4t_2^2\sin^2 k_y\simeq (4t_1^2\sin^2 \tilde{k}_y)/\gamma$. For the three-dimensional band dispersion, Eq.~(\ref{eq:3D_1}), we can obtain
\begin{align}
\mathrm{PF}^{\mathrm{3D}} _{\mathrm{max};xx} \simeq \gamma \mathrm{PF}^{\mathrm{3D, iso}}_{\mathrm{max};xx}, \\
\mathrm{PF}^{\mathrm{3D}} _{\mathrm{max};yy} = \mathrm{PF}^{\mathrm{3D}} _{\mathrm{max};zz}\simeq \mathrm{PF}^{\mathrm{3D, iso}}_{\mathrm{max};yy} = \mathrm{PF}^{\mathrm{3D, iso}}_{\mathrm{max};zz}.
\end{align}

Next, we move on to the problem with degenerate two-dimensional anisotropic bands: Eq.~(\ref{eq:2D_2}).
Based on the discussion above, we can immediately represent the transport coefficients with degeneracy $K^{\mathrm{2D, degen}} _{xx;\nu}$ as follows:
\begin{equation}
K^{\mathrm{2D, degen}} _{xx;\nu}[\mu] \simeq (\sqrt{\gamma}+\frac{1}{\sqrt{\gamma}})K^{\mathrm{2D, iso}}_{xx;\nu}[\mu+2(t_2-t_1)],
\end{equation}
which yields Eq.~(\ref{eq:apr2}).
As for the three-dimensional band dispersions, Eq.~(\ref{eq:3D_2}), the similar relation, Eq.~(\ref{eq:apr4}), clearly holds.
Therefore, the PF enhancement factors $\alpha$ defined in Eq.~(\ref{eq:alpha}) obeys Eq.~(\ref{eq:apr5}): 
$\alpha_{\mathrm{2D}}\simeq \alpha_{\mathrm{3D}} \simeq 1+ \gamma^{-1}$.
For example, the isotropic band dispersion ($\gamma=1$) yields a power factor doubled by the two-fold degeneracy.
The relations shown in this appendix hold only when the assumption we used, i.e., $k_B T \ll t_1,t_2$, is valid. Therefore, the $t_2 \to 0$ limit that corresponds to $\gamma \to 0$ is out of the range for application.


\begin{thebibliography}{999}
\bibitem{note_BiCuSeO} The optimized crystal structure of BiCuSeO yields an artificial gap closing by the well-known underestimation of the band gap by PBE-GGA. Such gap closing happens to be avoided by using the experimental crystal structure, which is why we adopted it for BiCuSeO. 
We note that, if we use the modified Becke--Johnson (mBJ) potential~\cite{mBJ1,mBJ2}, which is known to give a reasonable band gap,
the artificial gap closing does not take place and a band structure similar to the one reported in this paper can be obtained both for the experimental and optimized crystal structures.
\bibitem{footnote_inplane} Please note that even in the monolayer CuS, there are `out-of-plane' $p$-$p$ transfer integrals from the definition given here.
\bibitem{footnote_sym} The same consequence occurs even if one takes other distant $d_{xz/yz}$-$p_z$ hopping amplitudes into account. This is guaranteed by the mirror symmetry.
\bibitem{footnote_BiCuSeO} We verified that the first-principles band structure calculated with the mBJ potential, which is known to give a reasonable band gap, also exhibits a strong hybridization of the Bi states with the valence bands. This trend was also reported several times in the literature using various methodologies~\cite{BiCuSeO_band_deform1,BiCuSeO_band_deform2,BiCuSeO_band_deform3,BiCuSeO_band_deform4,BiCuSeO_boltz}.
\end{thebibliography}
\end{document}